\documentclass[conference,compsoc]{IEEEtran}

\ifCLASSOPTIONcompsoc

  \usepackage[nocompress]{cite}
\else

  \usepackage{cite}
\fi

\ifCLASSINFOpdf

\else

\fi

\usepackage{tabularx}
\usepackage{silence}
\usepackage{url}
\usepackage{placeins}
\usepackage{xurl} 
\usepackage{tikz}
\usepackage[T1]{fontenc}
\usepackage{hyphenat}
\usepackage[utf8]{inputenc}
\usepackage{graphicx}
\usepackage{subcaption}
\usepackage{tikz}
\usetikzlibrary{decorations.pathreplacing, arrows.meta} 
\usepackage[most]{tcolorbox}
\newcounter{takeawaycount}
\usepackage{booktabs}
\usepackage{array}
\usepackage{float}
\usepackage{algorithm}
\usepackage{algpseudocode}
\usepackage{microtype}
\usepackage{hyperref}
\newtcolorbox{takeaway}{
    colback=gray!5,
    colframe=black!75,
    fonttitle=\bfseries,
    title=Observation~\stepcounter{takeawaycount}\thetakeawaycount,
    arc=0mm,
    left=2pt,
    right=2pt,
    top=2pt,
    bottom=2pt,
    boxrule=0.5pt
}
\newcommand{\papertitle}{PAGE404}

\newcommand{\yicheng}[1]{}
\newcommand{\ravan}[1]{}
\newcommand{\nael}[1]{}
\newcommand{\circled}[1]{\textcircled{\small #1}}
\begin{document}

\title{\papertitle{}: Breaking GPU
Isolation Through Hardware-assisted Page Migration}

\author{\IEEEauthorblockN{Ravan Nazaraliyev\IEEEauthorrefmark{1}, Yicheng Zhang\IEEEauthorrefmark{2}, and Nael Abu-Ghazaleh\IEEEauthorrefmark{1}}
\IEEEauthorblockA{\IEEEauthorrefmark{1}Department of Computer Science and Engineering, University of California, Riverside\\
Email: rnaza005@ucr.edu, naelag@ucr.edu}
\IEEEauthorblockA{\IEEEauthorrefmark{2}Department of Electrical and Computer Engineering, George Mason University\\
Email: yzhang95@gmu.edu\\[0.5em]}
\IEEEauthorblockA{\vspace{-1em}\textit{Accepted to the 48th IEEE Symposium on Security and Privacy (S\&P 2027)}}}

\maketitle

\begin{abstract}
NVIDIA's Grace Hopper (GH200) superchip represents a new generation of
heterogeneous computing, tightly integrating a 72-core Arm-based CPU with a
Hopper GPU on a single module via a high-bandwidth, cache-coherent
interconnect. By eliminating the conventional PCIe bottleneck that has long
constrained CPU--GPU data movement, this design establishes a unified memory
domain that scales to the capacities demanded by generative AI and
high-performance computing (HPC) workloads. The architecture also introduces
hardware-assisted page migration support: driven by access counters, the
system dynamically promotes frequently accessed pages from host CPU memory to the GPU DRAM to improve memory locality. In this paper, we study the security
of this emerging architecture against new types of side- and covert-channel
attacks targeting its page migration support. We analyze counter-driven
migration across both the hardware mechanisms and the UVM driver's
interrupt-handling pipeline, and identify two side-channel vulnerabilities: one
rooted in the shared structure of access counters, and the other in the
driver's response to migration interrupts. We further show that both channels are exploitable across Multi-Instance GPU (MIG) partitions, a feature intended to provide strong, hardware-enforced isolation between co-located tenants.
Using proof-of-concept attacks, we build covert channels that reach roughly
8{,}500 bits per second through eviction-based signaling and 8{,}100 bits per second through migration-delay-based signaling across MIG instances. The interrupt-based channel also fingerprints inference workloads across distinct large language models (LLMs) and CPU SPEC2017 benchmarks, and estimates the prompt length processed during LLM inference with high accuracy. Taken together, these results expose a security gap in hardware-managed heterogeneous memory systems and call into question the isolation guarantees of current
MIG-enabled multi-tenant GPU deployments. The artifact of PAGE404 is available at \href{https://github.com/RavanN700/PAGE404_AttacksOnGH200.git}{PAGE404 GitHub}.
\end{abstract}

\IEEEpeerreviewmaketitle

\maketitle

\section{Introduction}
\label{sec:intro}

The growing demands of machine learning models and high-performance computing~(HPC) workloads challenge the capabilities of modern CPUs, GPUs, and their memory systems.   Successive GPU generations have been developed with increasing computational and memory capabilities in an effort to meet these needs.  A significant source of overhead arises from memory management across the GPU and CPU: existing memory management approaches such as \textit{Unified Virtual Memory} (UVM) require expensive software-managed fault handling via the UVM driver to transfer pages from/to the CPU and GPU memories.  With the large memory demands of large models (e.g., to hold the KV cache), efficient integration between the GPU and CPU is needed to allow the GPU to use the host CPU memory.

To address these demands, newer designs, such as NVIDIA's Grace Hopper (GH200) superchip, support tighter CPU--GPU integration and hardware-managed memory systems to enable low-overhead uniform memory access between the GPU and CPU.  The GH200 integrates an ARM-based Grace CPU with a Hopper GPU via NVLink-C2C, a
cache-coherent, high-bandwidth interconnect supporting the Address Translation
Service~(ATS)~\cite{nvidia2022gracehopper}. ATS enables GPU threads to access
system-allocated memory directly through hardware-assisted virtual-to-physical address translation, eliminating most of the software managed page-fault overhead of conventional UVM.  The GH200 further introduces a hardware-assisted page migration policy that tracks page usage and automatically migrates hot pages to the GPU.  This policy is implemented through \textit{access counters}: per-region hardware counters that track GPU accesses to system memory pages and notify the UVM driver once a configurable threshold is reached.  The driver then migrates the corresponding pages from host DRAM to GPU-local memory,
optimizing memory locality without requiring programmer intervention. 
While this mechanism is designed for performance, shared access counters and the driver's interrupt-handling pipeline for servicing these notifications are subject to contention from co-located GPU workloads, causing observable migration delays related to other system activity.
 
In this work, we investigate whether these new mechanisms for hardware-assisted page-migration expose side channel and covert channel vulnerabilities. Because these platforms are expensive, they are frequently shared among potentially mistrusting users (e.g., in a cloud deployment).   We reverse-engineer the access-counter subsystem and the UVM driver's interrupt-handling mechanism, characterizing the structure and behavior of the page migration counters and how migration notifications are processed by the driver. From this analysis, we identify two previously unexplored side-channel vulnerabilities.

The first channel stems from the shared structure of the access-counter hardware across tenants. We discover that the counter hardware is organized into a set-associative structure similar to caches, and notifications are triggered deterministically once a migration threshold is reached.  Since this structure is shared, it can be exploited using contention (e.g., evicting other processes' counters) for covert communication across tenants.  Similarly, the leakage through these shared counters can be measured (observed through the presence or absence of migration) to identify victim accesses to the same counter set. The second channel exploits migration contention on the shared driver: calls to the driver (e.g., to launch a new kernel) can be detected since they affect the response latency to migration events.  
Crucially, both channels target a GPU-wide shared resource, which makes them exploitable \textit{across Multi-Instance GPU~(MIG) partitions}, breaking the isolation guarantees of MIG.  On GH200, MIG partitions are shared across two containers, which is currently the only mechanism by which MIG instances are exposed to users.

 Building on these findings, we demonstrate several covert- and side-channel attacks. The eviction-based covert channel achieves 8500\,bps at a 1.5\% error rate, rising to 15500\,bps at roughly 5\% error with two eviction sets, while the eviction-based side channel distinguishes among 10 different CUDA workloads with 94\% accuracy when they use system-allocated memory. The migration-delay leakage enables a covert channel of 8100\,bps at a 4\% error rate and, across several side-channel attacks, distinguishes CUDA and LLM workloads with around or above 90\% accuracy, estimating an LLM's processed token length with a mean absolute error of 4.6 tokens on average. Moreover, we show that even when a tenant runs no GPU workload at all, a spy can still infer side-channel information about that tenant's CPU workloads.  
 This is because page migration requires the writer-prioritized mmap\_lock to read the process's memory map. As a result, concurrent CPU memory operations (for example,  munmap calls) can delay the migration's read request.

\papertitle{} makes the following contributions:
\begin{itemize}
\setlength{\itemsep}{0in}
    \item We reverse-engineer the complex hardware-assisted page migration support present on the new NVIDIA GH200 superchip, characterizing its underlying operation and discovering two new leakage vectors.

    \item We design covert channel attacks that enable information exchange across isolated tenants operating on separate Multi-Instance GPU (MIG) partitions.
    \item We demonstrate a cross-tenant side-channel attack between processes located on different MIG partitions that exploits two distinct leakage vectors to spy on different workloads. Specifically, we accurately fingerprint a large language model (LLM) architecture and infer some information about the prompt executed by a victim workload.
    
\end{itemize}

\noindent \textbf{Responsible Disclosure.} We disclosed our findings to NVIDIA. 
In their response, NVIDIA acknowledged our findings and indicated that they do not consider this a security issue, stating that the observed side-channel leakage is coarse-grained and noting that NVLink is not shared across untrusted tenants~\footnote{We note that this rationale overlooks key aspects of our threat model: \textsc{PAGE404} does not require physical NVLink interconnect sharing between victim and attacker tenants. Furthermore, in multi-GH200 configurations, cross-tenant driver-level interaction and concurrent memory management operations can induce contention and observable migration side effects even without shared physical interconnects.}.

\section{Background and Threat Model}
\label{sec:background_threat_model}
This section overviews the NVIDIA memory management and outlines the threat model and attack assumptions.

\subsection{Background}
\label{subsec:background}

Memory management across heterogeneous processors such as CPUs and GPUs provides a unified abstraction over what are physically distributed memory resources (separate GPU and CPU memories), allowing automated management of memory without expensive run-times or burdensome programmer management. NVIDIA's Unified Virtual Memory~(UVM)~\cite{uvm} was an early step in this
direction, allowing CPU and GPU to share a single virtual address space. Heterogeneous Memory Management~(HMM)~\cite{hmm}, a later model, enables GPU threads to access system-allocated memory directly via shared virtual addressing. 
The overheads of memory management continue to be high, leading to more recent architectures, such as the Grace Hopper GH200 superchip, that incorporate hardware-assisted address translation and dynamic page migration~\cite{NVIDIA2025PageMigration,
nvidia2022gracehopper} to improve performance and to provide coherence across an address space that spans the CPU and GPU memories. 

Workloads whose memory footprint exceeds GPU capacity can benefit from this tight integration. Consider the key-value (KV) cache in large language model inference: as sequence lengths and batch sizes grow, the KV cache can quickly outgrow the available GPU memory~\cite{pan2024instinfer}, forcing developers to either cap context length or manually stage data between host and device. Large-scale graph processing poses a similar challenge, where the graph's vertex and edge data may be far larger than GPU memory~\cite{gera2020traversing, nazaraliyev2025dream}, yet access patterns are irregular and data-dependent, making static partitioning ineffective. In both cases, the programmer is traditionally burdened with explicit memory management, allocating device buffers, orchestrating host-to-device transfers, and tracking which data resides where. This motivates the critical primitives these systems aim to provide: a unified address space so CPU and GPU can reference the same data through a single pointer, on-demand paging so data migrates automatically only when accessed, and transparent oversubscription so the working set can exceed physical GPU memory without manual intervention. The systems described below address these needs to varying degrees.

\begin{figure}
    \centering
    \includegraphics[width=1\linewidth]{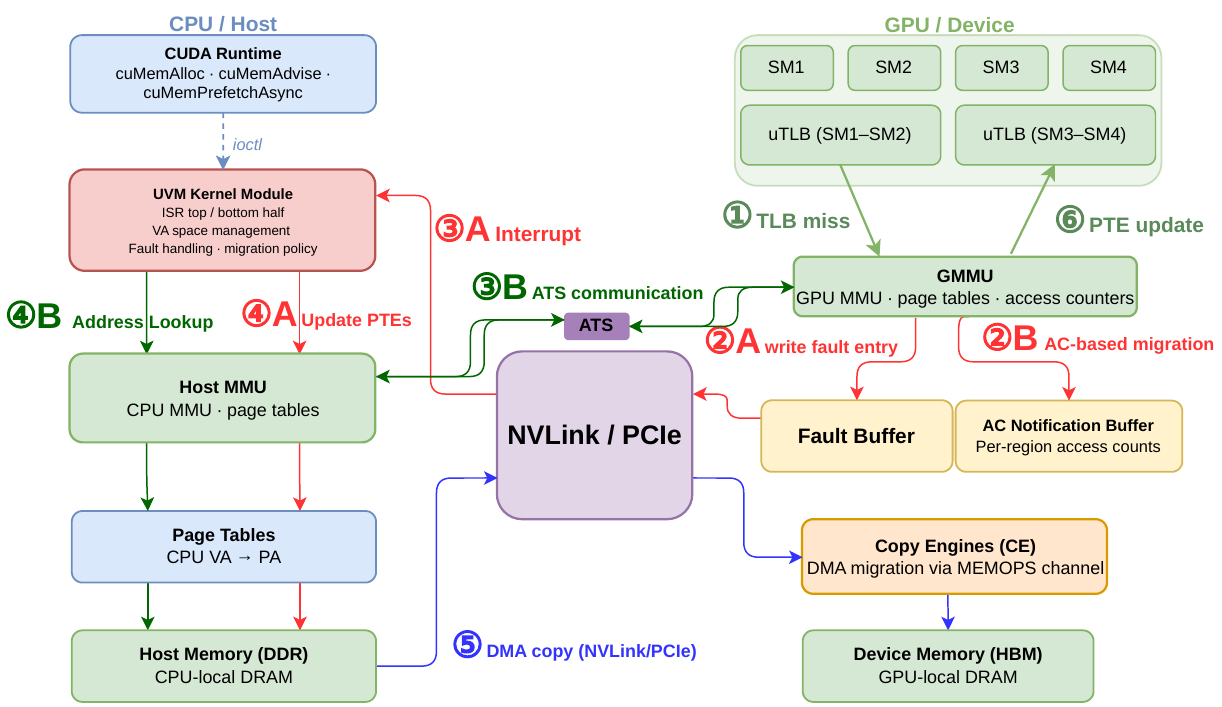}
    \caption{Unified Virtual Memory diagram.}
    \label{fig:uvm_diagram}
\end{figure}

\noindent\textbf{UVM Demand-Fault Migration.}
Figure~\ref{fig:uvm_diagram} illustrates a high-level overview of address translation and
migration paths present in UVM implementations. On the CPU side, the CUDA Runtime exposes a programmer-facing API for memory allocation and 
it communicates with the Kernel Module via \texttt{ioctl} system calls, which manage the virtual-address~(VA) space mapping and migration policies. On the GPU side, Streaming
Multiprocessors~(SMs) each support their address translations through a micro-TLB~(uTLB), which
are managed by the GPU Memory Management Unit~(GMMU) responsible for maintaining the GPU page
tables. When a GPU thread accesses a page not currently mapped in its uTLB, a TLB miss is
raised~\circled{1} and forwarded to the GMMU. The GMMU writes a structured fault entry into the
Fault Buffer: a GPU-visible ring buffer shared with the CPU~\circled{2}A. 
Subsequently, it raises an
interrupt to the host~\circled{3}A. The UVM Kernel Module services this interrupt in its bottom
half, draining the Fault Buffer. The module then updates the CPU-side page tables to reflect the
new page placement~\circled{4}A, and instructs the Copy Engines~(CE) to perform a direct memory
transfer of the faulting page over NVLink or PCIe from Host Memory to Device Memory
(HBM)~\circled{5}. Once the copy completes, the GMMU updates its own page table entries and
invalidates the relevant uTLB entries so that the faulting instruction can be replayed
transparently~\circled{6}.

\textbf{Heterogeneous Memory Management (HMM).}  HMM extends UVM's shared virtual addressing to any system-allocated memory (ordinary \texttt{malloc}, \texttt{mmap}, etc.) by integrating with the OS page-fault subsystem, whereas UVM only works for memory explicitly allocated through \texttt{cudaMallocManaged}.   While this may improve the programmer
productivity, it leads to substantial overheads, which motivated the hardware-assisted memory management support present in the Grace Hopper superchip.

\textbf{Hardware-assisted memory sharing.} GPU memory accesses may trigger page faults that require the UVM/HMM kernel module to intervene, introducing interrupt-handling overheads and additional OS interactions that can substantially impact performance. UVM/HMM also lacks hardware cache coherence between the CPU and GPU, requiring developers to manage memory coherence manually or limit updates from both the CPU and the GPU.  Given that optimal page placement is often unpredictable and may change at run time as access patterns change, static page placement is difficult. 
\begin{figure}
    \centering
    \includegraphics[width=1\linewidth]{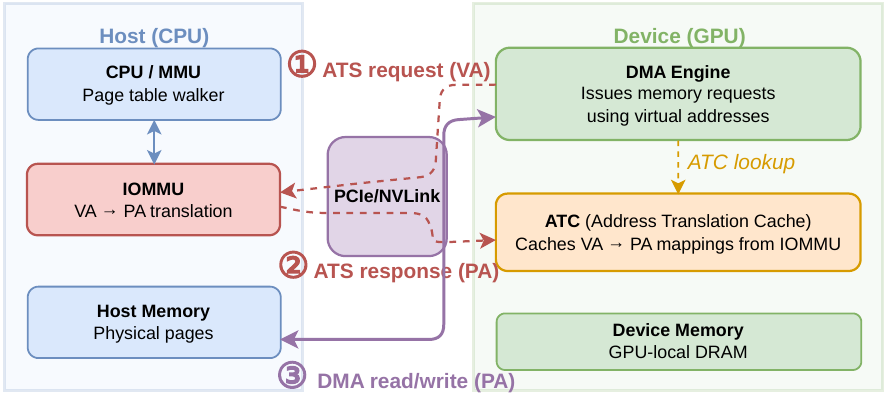}
    \caption{Address Translation Service in PCIe interconnect.}
    \label{fig:ats_diagram}
\end{figure}

As shown in Figure~\ref{fig:ats_diagram}, the host side comprises the CPU and its Memory Management Unit~(MMU), which owns the page tables, and an Input/Output MMU~(IOMMU) that enforces address translation for device-initiated memory accesses. On the device side, the DMA Engine issues memory requests using virtual addresses and is backed by an on-chip Address Translation Cache~(ATC) that stores recently resolved VA$\to$PA mappings. 
When the DMA Engine needs to access a host memory location, it first checks its local ATC. On a miss, it sends an \textit{ATS translation request}~\cite{Arm_SMMU_Guide, Intel_Arria10_ATS} carrying the virtual address across the PCIe link to the host Root Complex, which forwards it to the IOMMU~\protect\circled{1}. The IOMMU walks the page tables, resolves the physical address, and returns an \textit{ATS
translation response}~\protect\circled{2}; the device caches this VA$\to$PA mapping in its ATC
for subsequent accesses. With a valid translation cached, the DMA Engine proceeds to issue
physical-address DMA read/write operations directly to host memory over the PCIe
link~\protect\circled{3}, bypassing the need for further translation requests. On the GH200,
NVLink-C2C~\cite{nvlinkc2c} extends this mechanism with cache coherence, enabling the GPU to access host memory without explicit page migration. Within UVM, this serves as the hardware-assisted alternative to the fault-and-copy
path in Figure~\ref{fig:uvm_diagram} (\protect\circled{3}B--\protect\circled{6}). 

\begin{figure}
    \centering
    \includegraphics[width=1\linewidth]{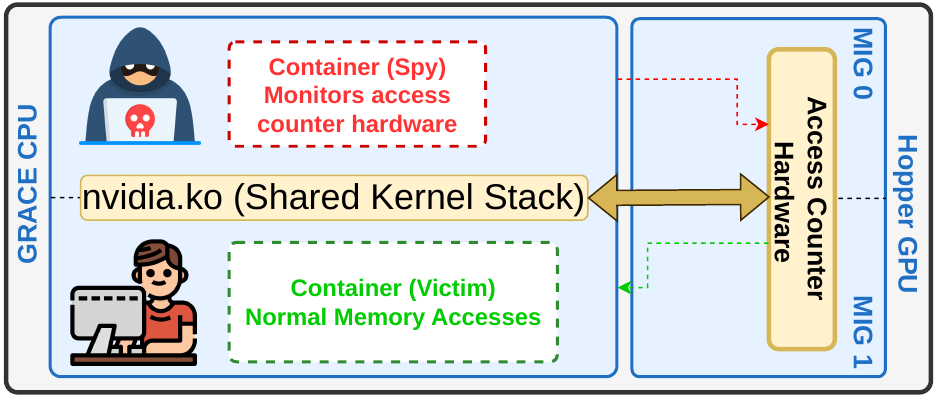}
    \caption{Threat model.}
    \label{fig:threat_model}
\end{figure}

\noindent\textbf{Access-Counter-Driven Migration.}
In parallel, the GMMU maintains hardware \textit{access counters} that track how frequently the memory region is accessed remotely from the GPU. Once a configurable access threshold (default 256) is exceeded, the GMMU writes a notification entry into the AC Notification Buffer and asserts an interrupt (``counter~$\geq$~threshold'')~\circled{2}B. The UVM Kernel Module
receives this as an \textit{AC notification} and, rather than waiting for a hard fault,
proactively issues a \textit{GPFIFO push} command to the Copy Engines, triggering a migration of the frequently accessed pages to Device Memory~\circled{5}. Unlike the demand-fault path, this migration does not stall the executing GPU thread; it is treated as a placement hint rather than a mandatory demand. The access counter structure is shared among MIG instances, and migration is handled in the driver kernel, shared among tenants, together with primary sources of contention analyzed in this work.

\subsection{Threat Model}
\label{subsec:threat_model}
\papertitle{} assumes the highest level of isolation between tenants that use different Multi-Instance GPU slices. MIGs are allocated to users in containers~\cite{nvidia_container_toolkit}, and Kubernetes Engines can be utilized to handle the sharing mechanism~\cite{gke}. NVIDIA also provides vGPU technology~\cite{nvidia_vgpu}, where each user has its own guest driver, but it is not supported on ARM platforms. Hence, the highest isolation level is that each user gets its own MIG slice within their container. Containers share the same GPU driver (nvidia.ko in Fig.~\ref{fig:threat_model}).
Previous works assumed sharing of a GPU with other tenants~\cite{naghibijouybari2018rendered, wei2020leaky, zhang2024invalidate+, lee2014stealing, zhang2023t}, where a \textit{spy} CUDA application and a \textit{victim} GPU context, i.e., a CUDA or graphics application, share a GPU. Other works assumed GPU sharing in cloud settings, where multiple tenants can share the same physical GPU through virtualization~\cite{tan2022gpupool, diab2013dynamic, li2011gpu, xue2018scalable}. Additionally, prior research has demonstrated practical techniques for achieving co-location in cloud systems~\cite{atya2017malicious, zhao2024everywhere, ristenpart2009hey}. MIG attack papers~\cite{zhang2023t, gubehind} assumed that different users get their own MIG instance. The attacker does not require superuser privileges. Instead, they perform experiments to reverse-engineer the page migration details and the behaviour of hardware access counters, as a preliminary step for the attacks. 
\section{Demystifying HW-assisted Page Migration}
\label{sec:reverse}

This section describes the methodology we use to reverse-engineer undocumented aspects of page migration and the hardware access-counter mechanism on the GH200 Superchip. Experiments are performed on GH200 nodes from both
CloudLab~\cite{Duplyakin+:ATC19} and Lambda Cloud. The modified driver used
throughout this section is available in our artifact repository.

\begin{figure}
    \centering
    \includegraphics[width=1\linewidth]{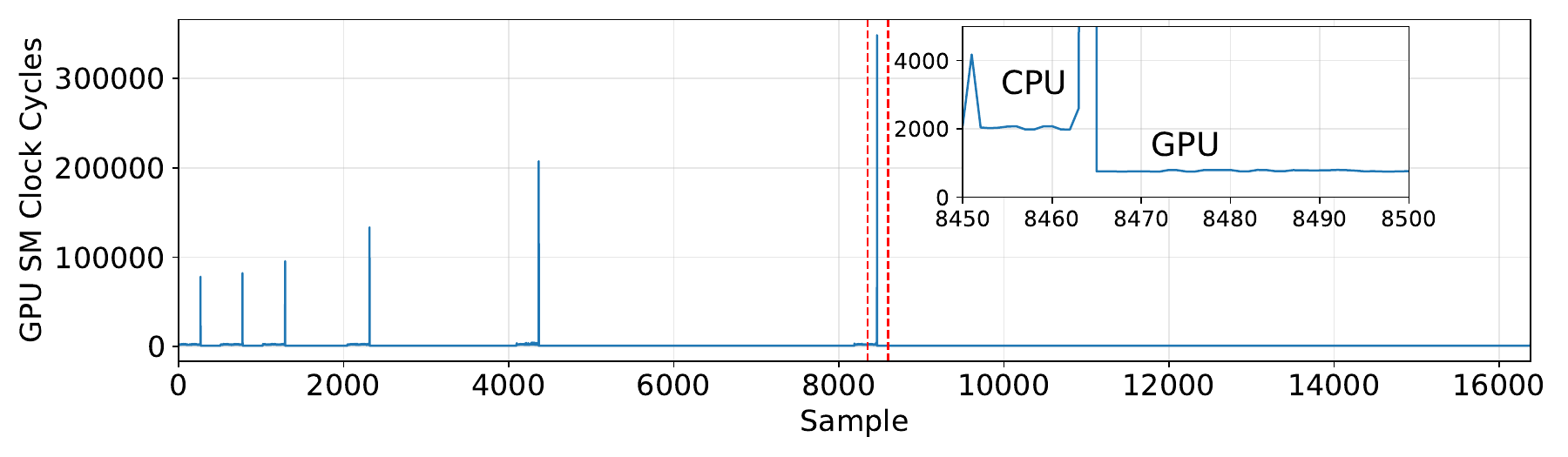}
    \caption{Hardware-assisted page migration on GH200.}
    \label{fig:page_mig}
\end{figure}
\subsection{Page Migration Access Counters}
\label{subsec:how_to_count}

\begin{figure}
    \centering
    \includegraphics[width=1\linewidth]{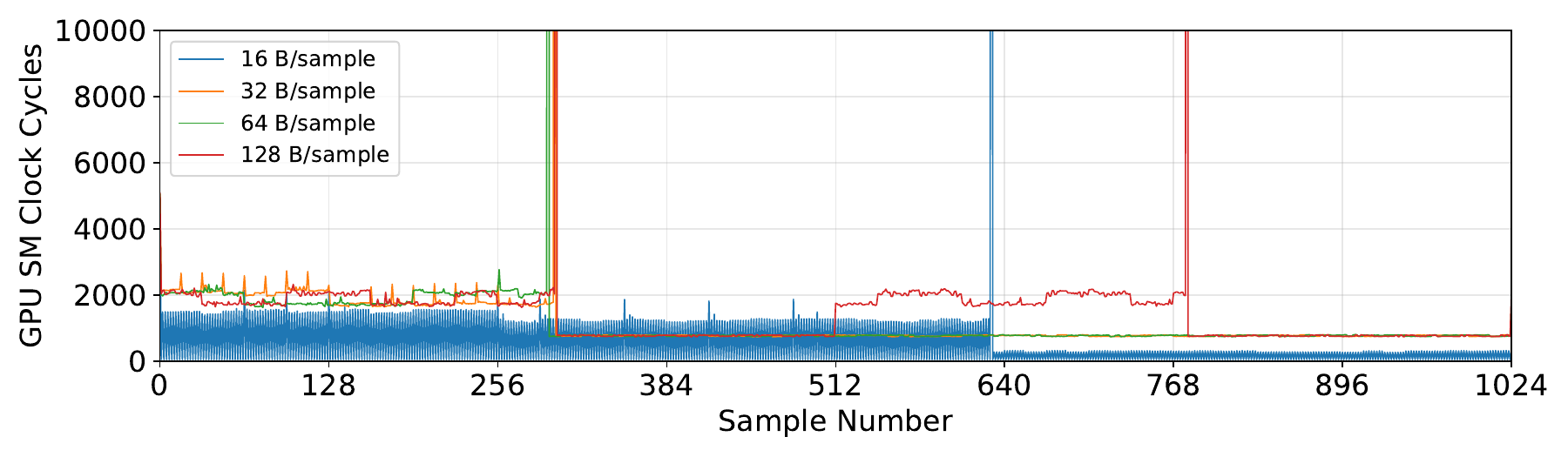}
    \caption{Varying size of the stride.}
    \label{fig:stride_size}
\end{figure}
\textbf{Overview of the page migration.} To understand the behavior of page migration, we design an
experiment that performs strided memory accesses over a system-allocated (i.e., CPU allocated) memory
region using a single GPU thread. The resulting access latencies reveal a
repeatable migration pattern that recurs over every $2\,\text{MB}$ region.
Figure~\ref{fig:page_mig} depicts a representative $2\,\text{MB}$ virtual
address space, exposing the underlying migration mechanism. The migration behavior is
explained by two complementary mechanisms: page migration driven by access
counters, and prefetching performed by the UVM driver module. Prefetching is
governed by a tree-based promotion policy implemented within the UVM driver.
Migration operates at the granularity of a single $64\,\text{KB}$ page: once the
access counter for a given $64\,\text{KB}$ page reaches a specific threshold, that page is migrated to GPU memory. With a sequential access pattern, upon each migration, potentially additional regions are
prefetched corresponding to one node higher in the promotion tree, so that the
number of prefetched $64\,\text{KB}$ pages doubles at every migration event,
following the sequence $1$--$1$--$2$--$4$--$8$--$16$ pages until the full
$2\,\text{MB}$ region has been promoted to HBM.  Under a random access page migration pattern, a subtree is prefetched if more than half its leaf 64KB pages have already been migrated.

\begin{takeaway}
    Hardware-assisted page migration operates at a $64\,\text{KB}$ page
    granularity. The prefetching scheme is the traditional tree-based UVM
    prefetcher, and prefetching occurs within a single $2\,\text{MB}$ virtual
    page.
\end{takeaway}

\textbf{Determining the cacheline size. }In this experiment, we vary the stride size to determine the cache line size of GPU memory accesses. Figure~\ref{fig:stride_size} shows access latencies for four different stride sizes. For stride sizes of 128, 64, and 32 bytes, the number of accesses required to trigger page migration remains constant at 256. However, when the stride size is reduced to 16 bytes, the number of accesses required to trigger page migration doubles. Furthermore, at a stride size of 16 bytes, the access latency alternates between that of CPU main memory (DDR) and GPU L1 cache, indicating that two consecutive 16-byte accesses fall within the same 32-byte cache line---with the first access fetching the line from DDR and the second being served from the L1 cache. This observation is consistent with prior work~\cite{nazaraliyevnot,zhang2025nvbleed,zhang2024invalidate+} and confirms that the GPU cache line size for system memory accesses is 32 bytes~\cite{nvidia_memory_transactions}. Finally, we repeat the experiment using \texttt{load} instructions annotated with the \texttt{volatile} keyword and observe that data is not cached at any memory hierarchy level when the GPU accesses system memory through \texttt{volatile} loads, bypassing the cache entirely.

\begin{takeaway}
    The cacheline size is 32 bytes. The load operation used to access a \textit{volatile} memory does not cache the data at any GPU cache level.
\end{takeaway}

\begin{figure}
    \centering
    \includegraphics[width=1\linewidth]{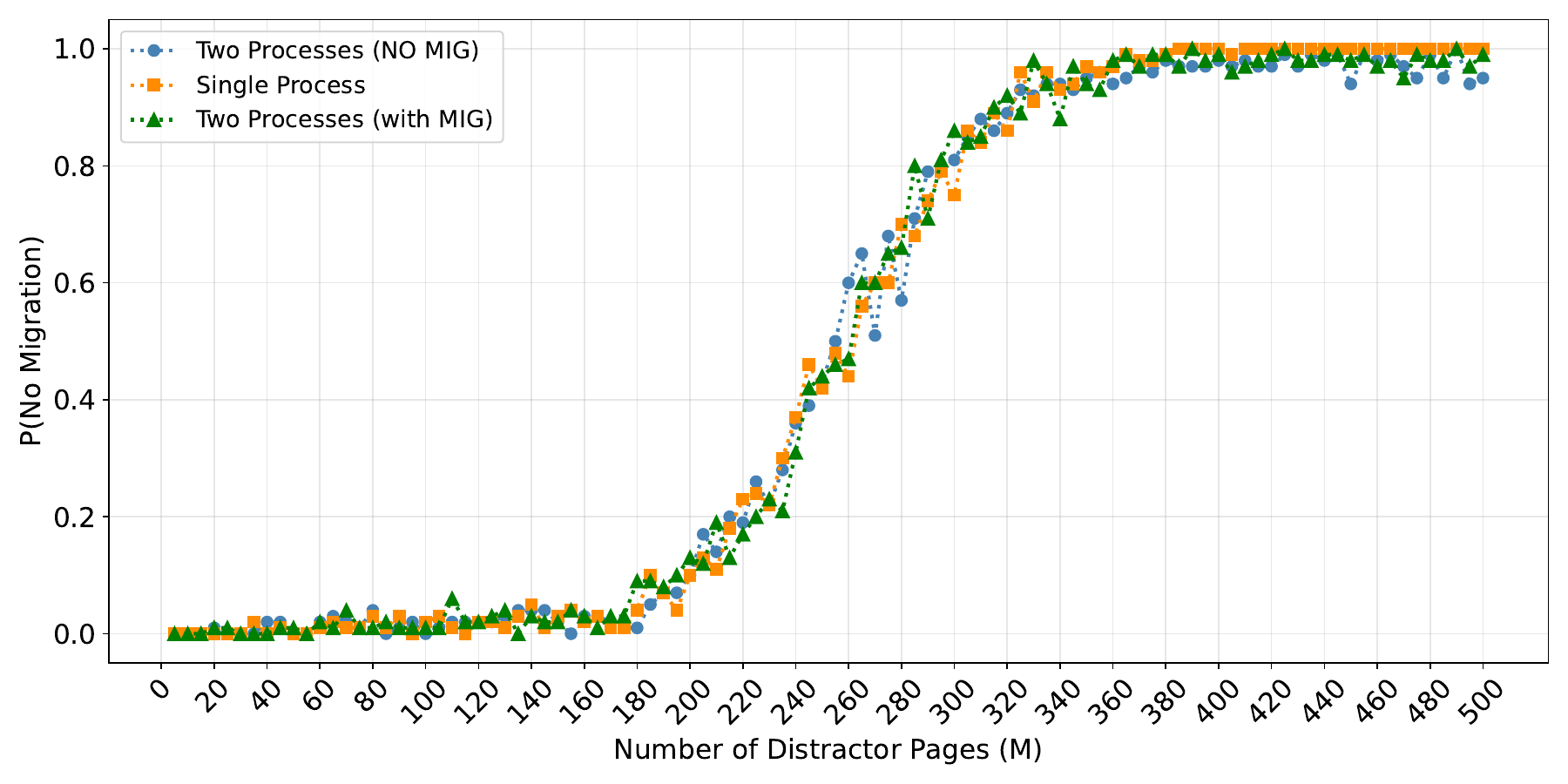}
    \caption{Probability of no migration of first page when M distractor pages are accessed in Phase 2.}
    \vspace{-0.5cm}
    \label{fig:num_counters}
\end{figure}

\textbf{Hardware Access Counter Circuit and Sharing Mechanism.} As described in~\cite{NVIDIA2019MultiGPU, NVIDIA2025PageMigration}, hardware counters in GH200 are implemented as a finite pool, where each counter is assigned to a specific address range, either virtual or physical, according to a replacement policy such as Least Recently Used (LRU)~\cite{NVIDIA2019MultiGPU}. In this experiment, we probe the size of this counter pool. The experiment proceeds as follows: a process first accesses $N - \delta$ addresses to bring a target page's counter close to the migration threshold. Using the default threshold of $N = 256$ and setting $\delta = 2$, only two additional accesses are needed to trigger hardware-assisted page migration for the target page. We then introduce $M$ \textit{distractor} pages, each accessed $X < N$ times, forcing the pool to assign counters to these new pages. After making $X$ accesses to each of the $M$ distractor pages, we perform the remaining $\delta$ accesses to the original target page and observe whether migration is triggered~\footnote{Migration can be detected either by inspecting the page's physical address or by repeatedly accessing the page and determining whether it migrates within an acceptable number of memory accesses, a quantity we later denote as the migration delay.}. Figure~\ref{fig:num_counters} shows (Single process) the probability of hardware-assisted page migration as a function of $M$, the number of distractor pages. The results indicate that the counter pool is exhausted at approximately 360--380 distractor pages. The lack of a sharp cutoff in the figure is attributable to a behavior we characterize and explain later: distinct 64KB pages residing within the same 2MB page are assigned the same counter, producing an apparent eviction effect that is, in reality, a consequence of using the same counter.

\textbf{Hardware Access Counter Sharing across 2 Processes and MIG instances. } 
We extend the previous experiment to examine counter pool behavior across two independent processes, comparing execution on a full GH200 (without MIG partitioning) against execution on isolated Multi-Instance GPU (MIG) instances. The experimental setup is restructured into three phases: \textit{Phase 1}, in which Process~1 probes its target page with $N - \delta$ accesses and then blocks awaiting a signal; \textit{Phase 2}, in which Process~2 accesses its own $M$ distractor pages; and \textit{Phase 3}, in which Process~1 resumes and attempts to trigger migration on the target page. As shown in Figure~\ref{fig:num_counters}, the observed behavior is consistent across both the non-MIG and MIG configurations, with migration probability degrading at the same range of $M \approx 360$--$380$ distractor pages regardless of whether the two processes are isolated within separate MIG instances. This demonstrates that the hardware counter pool is shared across MIG boundaries, meaning that MIG's process isolation guarantees do not extend to the page migration hardware counters.

\begin{takeaway}
    The hardware counter pool is finite and is shared across Multi-Instance GPU (MIG) instances, violating the isolation guarantees expected of MIG partitioning.
\end{takeaway}

\textbf{Deep dive into Hardware Counters.} 
\noindent \textbf{Virtual or Physical addresses?} While we have shown that hardware counters are shared across processes and MIG instances, this does not reveal how counters are assigned. Although Figure~\ref{fig:num_counters} shows that the counter pool saturates at approximately 380 entries, migrations still occasionally fail even beyond 400 distractor pages, suggesting a probabilistic assignment mechanism. This probabilistic behavior arises because counters may be assigned based on either virtual addresses (VA) or physical addresses (PA)~\cite{NVIDIA2019MultiGPU}, both of which are randomized across process invocations due to ASLR and OS memory allocation.
To determine whether counter assignment is VA-based, we disable \texttt{ASLR}, ensuring that the process receives the same VA on every run. The experimental setup follows the previous experiment (single-process configuration) with this modification. If counters are assigned based on VA and the pool employs an LRU eviction policy, repeatedly running the process with fixed VAs should produce a deterministic eviction set of size $E$ beyond which migration consistently fails. However, we observe entirely probabilistic behavior even with \texttt{ASLR} disabled, indicating that VA plays no role in hardware counter assignment.

\begin{takeaway}
    Hardware access counters assignments are based on the physical addresses of pages.
\end{takeaway}

\noindent \textbf{Understanding Hardware Structure}. Since the counter pool is finite, understanding the counter assignment policy reduces to finding an \textit{eviction set}: the minimum number of distractor pages required to evict the counter assigned to a target page. This problem is analogous to cache eviction set construction, but introduces unique challenges: (1) once a page's counter reaches the migration threshold, the page is migrated to GPU DRAM and is no longer resident in host memory; and (2) migrating a page back to host memory does not preserve its physical address, as the original physical frame is released to the OS and a new allocation is made upon return. One way to handle this is to pin the host page; however, pinning disables the page migration.

\begin{figure}
    \centering
    \includegraphics[width=1\linewidth]{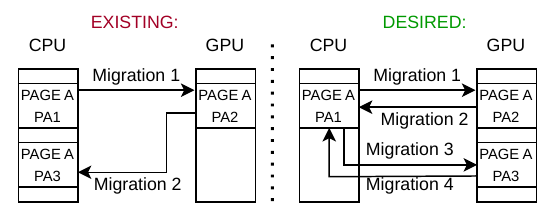}
    \caption{Memory migration mechanism and desired state.}
    \label{fig:reverse_challenge}
\end{figure}

The CUDA API \texttt{cudaMemPrefetchAsync()} can be used to prefetch the page back to host memory\footnote{\texttt{cudaCpuDeviceId} is passed as the destination device identifier.}. However, as illustrated in Figure~\ref{fig:reverse_challenge}, once a page is migrated to the device, its host-side physical frame is freed and returned to the OS. Prefetching the page back to host memory (Migration~2 in the existing flow) triggers a fresh allocation, assigning an arbitrary new physical frame. To overcome this limitation, we instrument the open-source NVIDIA driver to implement a modified page migration scheme: when a host-allocated page is migrated to the GPU, whether via hardware counters or any other mechanism, its physical frame is not released to the OS; instead, it is retained in a \textit{page cache}. Upon prefetching or migrating the page back to host memory, the cached physical frame is restored, ensuring physical address stability across migration events.
These modifications enable eviction set construction experiments analogous to those used in cache side-channel research. 

\begin{figure}
    \centering
    \includegraphics[width=1\linewidth]{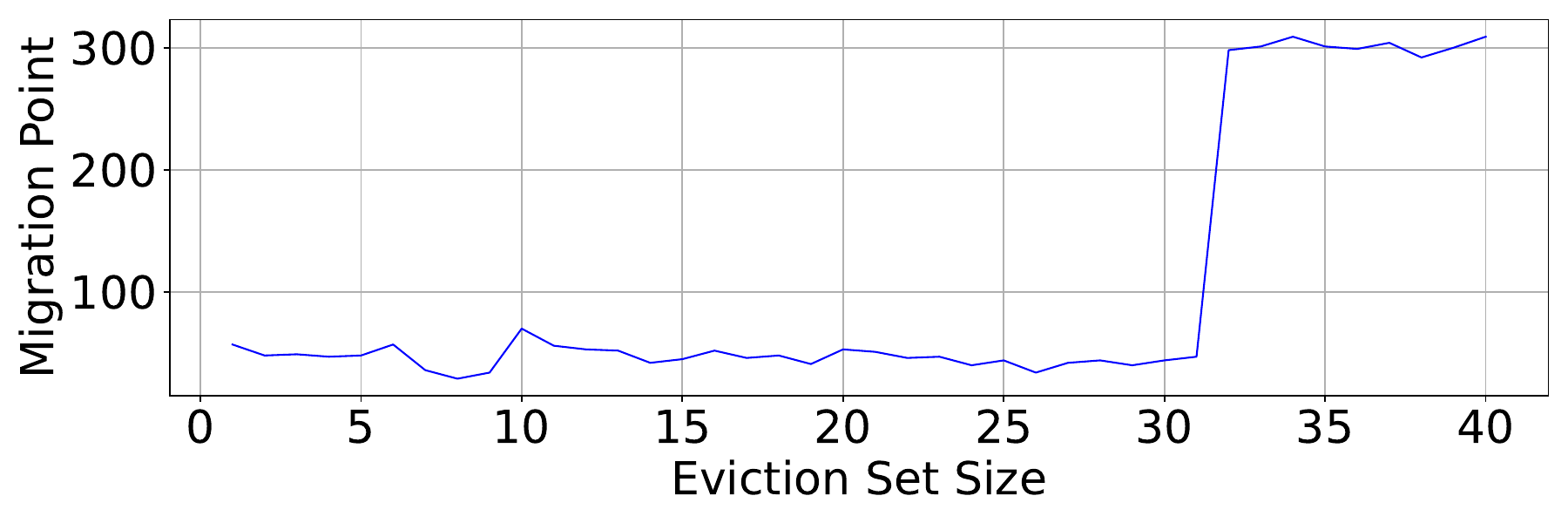}
    \caption{Hardware access counter set size. For illustration, we access the page continuously and observe when migration occurs; a value higher than 256 indicates that the counter had previously been evicted, so a fresh set of 256 accesses was required to trigger migration.}
    \label{fig:eviction_set_size}
\end{figure}

We construct the eviction set using the following procedure: the target page is first accessed $N - \delta$ times, bringing its counter close to the migration threshold; then $M$ distractor pages are accessed; finally, the target page is accessed again to determine whether migration is triggered, following the methodology of previous experiments. The driver modification is employed to reset counters and bring pages back to the host between trials: each page is accessed $N$ times to force its migration to GPU DRAM, after which it is prefetched back to host memory, restoring its physical frame and resetting its counter. We then determine the minimum number of distractor pages sufficient to consistently prevent the target page from migrating, which corresponds to the eviction set size. Additionally, by identifying the number of unique eviction sets, we derive the total number of hardware access counters. The results reveal that each counter set consists of 32 access counters (as illustrated in Fig.~\ref{fig:eviction_set_size}), with 8 unique sets, yielding a total of 256 hardware access counters across the GPU.

\begin{takeaway}
    The hardware access counters are distributed across 8 sets, with 32 counters in each set.
\end{takeaway}

\begin{figure}
    \centering
    \includegraphics[width=\linewidth]{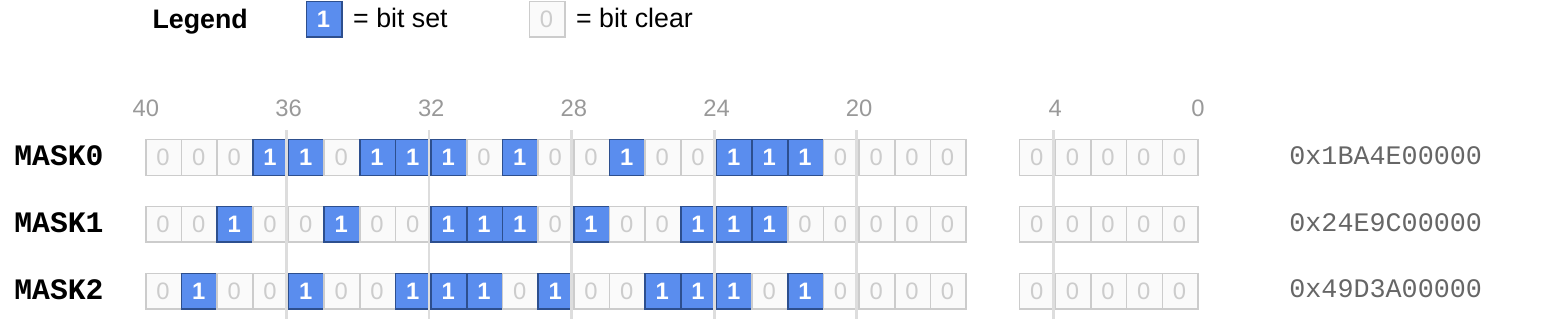}
    \caption{Bit-level view of each XOR mask (bit~39 = MSB $\rightarrow$ bit~0 = LSB).}
    \label{fig:xor_masks}
\end{figure}

\textbf{Understanding mapping of pages to counters.} 
Prior work has reverse-engineered the hash functions that map PAs to DRAM banks~\cite{pessl2016drama,jattke2024zenhammer}, or to LLC cache sets~\cite{kayaalp-16,liu2015last} under the assumption that these functions are linear (i.e., XOR-based). Adopting the same assumption, we search candidate XOR functions against the addresses we previously identified as belonging to the same set.  We use the functions to validate that our conflict sets are consistent.  We then compute the null space of the resulting system, and as a final validation step, we verify that the recovered functions produce distinct set IDs across the 8 groups of PAs. Using the modified driver, we collect 8 groups comprising approximately 4{,}700 addresses in total and run the brute-force search. The procedure recovers exactly 3 independent functions in the null space, matching the expected number of selector bits ($8 = 2^3$). The recovered masks are shown in Figure~\ref{fig:xor_masks}. All three functions operate exclusively on bits 21 through 38, which implies that every 2\,MB physically contiguous range is mapped to a single counter.

\textbf{Counter assignment vs.\ page migration.}
Because counters are assigned at a 2\,MB granularity, a discrepancy arises with the system page size, which also defines the default migration granularity of 64\,KB. We design an experiment to determine how access to different 64\,KB sub-regions within the same 2\,MB range contributes to migration decisions. Across multiple trials, we observe that accesses to \emph{any} 64\,KB sub-region within a 2\,MB range increment the same counter, and that migration is performed on the specific 64\,KB region targeted by the most recent access.

\begin{takeaway}
Each access counter is owned by a 2\,MB physical range, while migration is performed at the granularity of the most recently accessed 64\,KB region within that range.
\end{takeaway}

\begin{figure}
    \centering
    \includegraphics[width=1\linewidth]{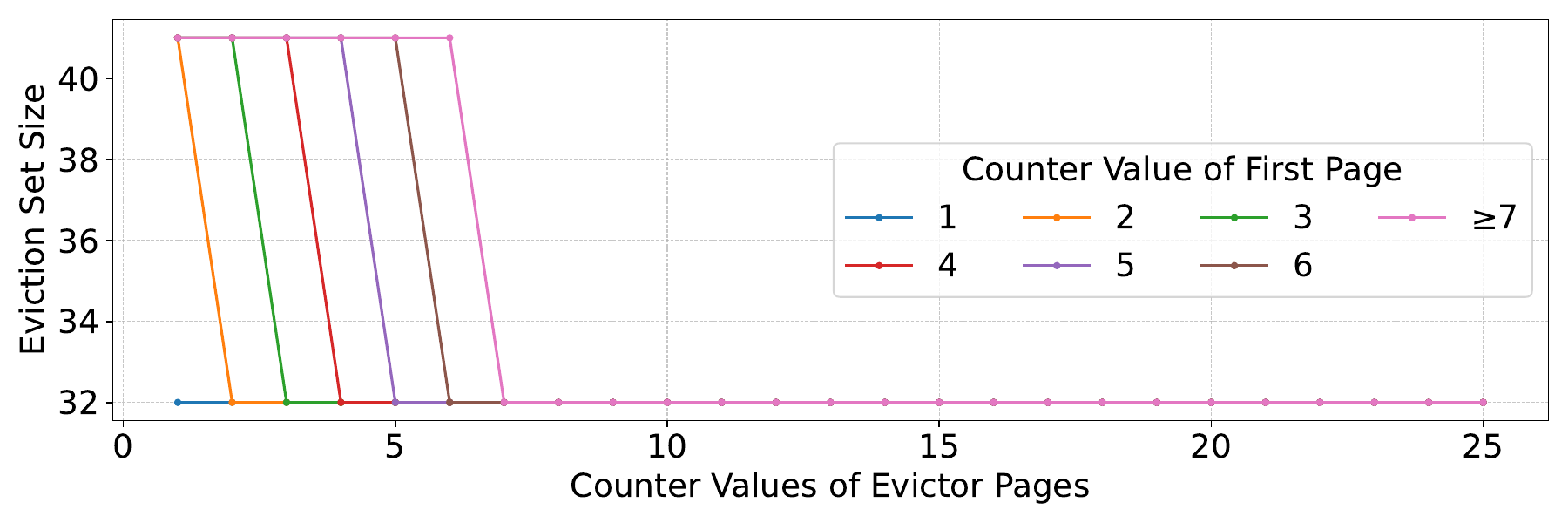}
    \caption{Effect of the counter value on the eviction policy. The eviction set size is computed as $40 - n_{\text{evict}} + 1$, where 40 denotes the total number of eviction elements used, $n_{\text{evict}}$ is the number of observed evictions, and the additive 1 accounts for inclusive counting.}
    \label{fig:eviction_set_sizes}
\end{figure}

\textbf{Effect of Counter Values on the Eviction Policy.}
Throughout our experimentation, we assumed a Least Recently Used (LRU)
eviction policy and held the counter values constant across all page accesses,
ensuring that our measurements were not confounded by variations in the counter
state. In this experiment, we relax that assumption to determine whether the
counter values themselves influence the eviction decision.
We first construct an eviction set of $40$ addresses, each residing in a
distinct $2\,\text{MB}$ region but mapping to the same counter set, following
the methodology established in our earlier findings. We then execute a pair of
nested loops: for a fixed number of accesses $n$ to the target page, we sweep
the number of accesses $m$ to each page in the eviction set. To obtain
noise-free measurements, we employ a second, independent eviction set to reset
the counter values in the counter set between trials. Each trial first issues $n$ accesses to
the target page and subsequently issues $m$ accesses to the eviction pages.
Figure~\ref{fig:eviction_set_sizes} reports the resulting behavior, where each
line corresponds to a distinct counter value of the target (first) page. As we
jointly vary $n$ and $m$, we observe that the counter values do in fact govern
the eviction decision. When a counter must be evicted, the victim is the counter
holding the smallest value, provided that value is below $7$. Once all counter
values reach or exceed $7$, the counter value ceases to influence the decision
and eviction reverts to LRU. The transition points visible in
Figure~\ref{fig:eviction_set_sizes}, where each curve drops from an eviction set
size of $41$ back to $32$, shift rightward as the target page's counter value
increases, confirming that a higher counter value confers stronger residency
until the saturation threshold of $7$ is reached. In summary, the access
counters prioritize eviction by counter value when that value is below $7$, and
otherwise fall back to LRU.

\begin{takeaway}
    The eviction policy evicts the counter with the smallest value when that
    value is below $7$, and reverts to LRU otherwise.
\end{takeaway}

\textbf{Counter Persistence Across Process Launches.}
Having characterized the eviction policy within a single execution, we investigate whether hardware access counter state persists across process boundaries. Specifically, once a counter tracks a physical address ($\mathrm{PA}$) during an initial execution, does that accumulated state survive when the same address is re-accessed in a subsequent process launch? 

To determine this, we first examine the GPU driver. Unified Virtual Memory (UVM) acquires the access-counter notification interrupt when the first virtual address space registers the GPU. This initialization path unconditionally resets the entire hardware counter table via \texttt{access\_counters\_take\_ownership()}, which dispatches a \texttt{MEM\_OP} command configured with \texttt{ACCESS\_COUNTER\_CLR} and \texttt{CLR\_TYPE = ALL}. Conversely, when the final address space unregisters upon exit, ownership is released without issuing a corresponding clear. Consequently, counter state is wiped only on the $0 \to 1$ transition of the driver's enablement reference count (i.e., when an incoming process initializes an otherwise idle GPU), rather than when the last process exits. A process launching onto a GPU that already hosts an active CUDA context therefore triggers no reset.

To empirically validate this mechanism, we evaluate page migration behavior across repeated launches with and without a background dummy GPU process. In each run, the target process issues 128 accesses to $\mathrm{PA}$ and terminates; upon relaunch, it issues an additional 128 accesses to the same address. When the GPU is initially idle, the page is not migrated, indicating that the counter resets to zero upon relaunch. In contrast, when a persistent dummy process holds the GPU context open, the subsequent launch triggers migration upon reaching the cumulative threshold of 256 accesses ($128 + 128$). These results demonstrate that counter state is decoupled from process lifecycle events such as context teardown and IOTLB invalidation, indexing purely by physical address and persisting until explicitly reset by driver initialization.

\begin{takeaway}
    Hardware access counters index strictly by physical address and are unperturbed by process teardown. However, counter state is unconditionally reinitialized to zero whenever a process initializes an otherwise idle GPU.
\end{takeaway}

\subsection{Migration: Driver}
\label{subsec:how_migration_happens}

\begin{figure}[t]
    \centering
    \includegraphics[width=1\linewidth]{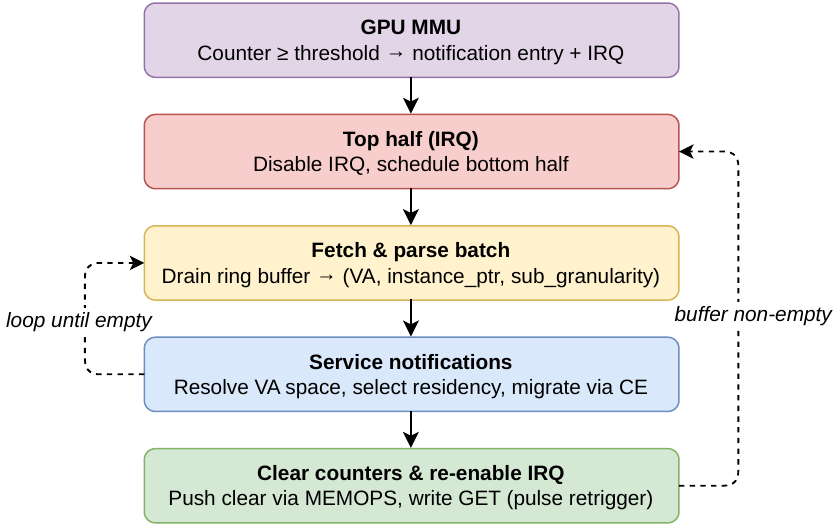}
    \caption{Sequential stages from the $N$th remote access to page residency,
             with the contention source at each stage.}
    \label{fig:contention_pipeline}
\end{figure}

The GPU's memory unit (GMMU) keeps a running count of how often each region of
memory is accessed remotely. Once that count crosses a set threshold (256 by
default), the hardware writes a notification into a ring buffer and raises an
interrupt. As shown in Figure~\ref{fig:contention_pipeline}, the driver handles
this interrupt in stages.

First, a lightweight \textit{top half} runs: it disables the interrupt and
schedules the heavier work for later. That heavier work, the \textit{bottom
half}, then drains the ring buffer in batches, parsing each notification into the
information needed to act on it (which address, which GPU, etc.). Next, in the \textit{service} stage, the driver figures out
which memory space each notification belongs to, picks where the data should
live, and issues a Copy Engine~(CE) migration to move it there. Finally, once the
copy is done, the driver clears the counters and re-enables the interrupt, which
prompts the hardware to check whether more notifications are waiting. If the
buffer still has entries, the cycle repeats; otherwise it loops back to wait for
the next batch.

Every one of these stages can add delay, and because they run strictly one after
another, those delays simply pile up into the total time between an access and
the data actually moving. But the bigger problem is \textit{variability}, and it
comes from sharing. The bottom half uses the same work queue as the GPU's
page-fault handlers, so if a batch of page faults is already in flight, the
migration has to wait for it to finish first. The situation gets worse because of
locking. Several common operations, such as allocating or freeing memory, can
hold locks that the migration also needs.

We define \textit{migration delay} $\Delta$ as the latency from the Nth remote access to a target page, which triggers the hardware notification, to the start of data transfer over NVLink. 
Although NVLink contention is possible and potentially measurable~\cite{zhang2025nvbleed}, in practice, we find that the software pipeline stages described above dominate the transfer delay, particularly the shared work queue and lock contention.
To isolate this software overhead, we measure migration delay in units of accesses rather than time, counting how many accesses to CPU memory are possible before the migration completes. 
The normal migration delay (with idle background) in units of system memory accesses is typically 20 to 30 . 
\begin{takeaway}
    Hardware-assisted page migration is subject to variable delays caused by interrupt handling in the shared GPU Driver, and this timing variability constitutes an information leakage vector across tenants in the same platform.
\end{takeaway}
\section{Proof-of-Concept Attacks}
\label{sec:attacks}
Throughout this section, we demonstrate how the findings from Section~\ref{sec:reverse} can be leveraged to mount proof-of-concept attacks on the GH200 platform. Unless otherwise stated, all experiments are performed on a GH200 node from
CloudLab~\cite{Duplyakin+:ATC19} running NVIDIA driver version 590.48.01 and
CUDA Toolkit 12.2.

\begin{figure}
    \centering
    \includegraphics[width=1\linewidth]{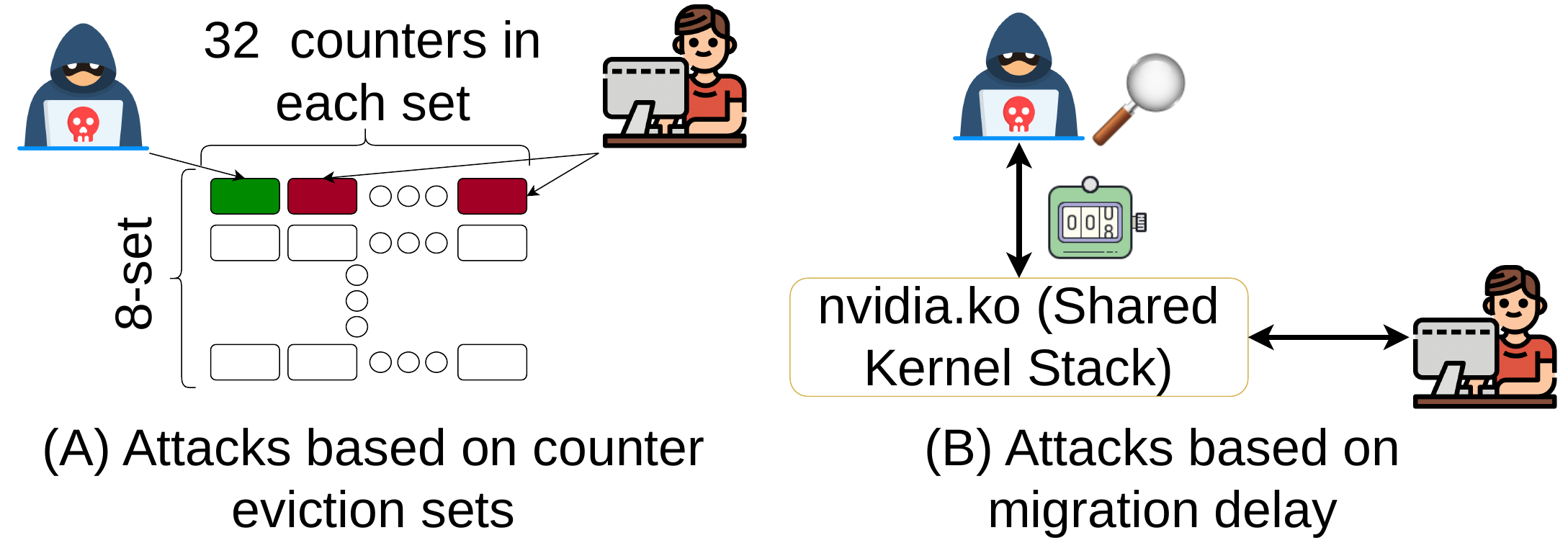}
    \caption{Attack scenarios}
    \label{fig:explanatory_fig}
\end{figure}

\subsection{Attack Overview}
\label{subsec:attack_overview}
We implement two types of attacks for both covert and side channel: (1) a channel based on counter eviction sets (Fig.~\ref{fig:explanatory_fig} A), and (2) a channel based on variation in the migration delay $\Delta$ induced by a co-located process (Fig.~\ref{fig:explanatory_fig} B). As demonstrated in \S\ref{sec:reverse}, both leakage vectors are exploitable across two processes running on separate MIG instances.

\subsection{Attack Challenges}
\label{subsec:attack_challenges}
Attacks share two core components: a \textit{probe phase} and a \textit{migration detection phase}. The critical parameter governing attack effectiveness is the sampling rate: a higher sampling rate increases the bit rate of the covert channel and enables the side channel to capture finer-grained patterns, ultimately leading to more accurate workload characterization and higher attack success rates. During the probe phase, the attacker performs $N$ accesses to the target page, incrementing the hardware counter toward the migration threshold. In the detection phase, the attacker observes if the migration happens.

\textbf{Speeding up Probing Phase.}  To speed up migration triggering, we experiment with 2 methods: N parallel accesses, or non-blocking loads. For parallel accesses, we can use $N=256$ parallel threads and only a single thread checks the migration delay. With non-blocking loads, we refer to loading data and not waiting for the result, so the code continues issuing load instructions. We measure the sampling rate of each method over 90GB of memory: around 17000 samples per second with N-parallel accesses and around 15000 samples per second with the non-blocking load method. Hence, parallel accesses are more suitable for capturing more details.

\textbf{Bypassing UVM Prefetching.} UVM operates on \emph{prefetch regions} of 2\,MB in the virtual address space. Within each region, page residency is tracked using a binary tree: the root represents the full 2\,MB, and each level splits its parent in half, down to a leaf granularity of 64\,KB. The driver uses this tree to detect locality; when one half of a node has been migrated, UVM treats the sibling half as likely to be accessed soon and prefetches it to GPU memory. This way, a few migrations in one part of a region trigger prefetching of the subsequent pages, reducing the number of migrations an attacker can exploit. To bypass this, we make the attacker operate at a 128\,KB stride: it touches one 64\,KB page, then skips the adjacent 64\,KB page before touching the next. By leaving each touched page's sibling leaf unoccupied, no node in the tree ever has one half fully resident, so prefetching is never triggered. This yields 16 migratable pages within a single 2\,MB region, while keeping each migration limited to a 64\,KB transfer with no prefetching triggered.

\textbf{Destructive Nature of Page Migration.}
Page migration is inherently destructive: once a page has been migrated to GPU memory, later accesses to it no longer fault, so the same page cannot trigger a migration again. PAGE404 handles this by moving to a fresh page after each migration. This step-forward approach also fits the UVM prefetching bypass discussed earlier: rather than advancing to the adjacent 64\,KB page, the attacker moves to a new 128\,KB-aligned region, so the next page is not covered by the prefetcher.

\subsection{Covert Channel Attacks}
\label{subsec:covert}
Both covert channels exploit contention, but through distinct mechanisms, one in hardware counting, the other in interrupt servicing. The first channel is entirely embedded in hardware, leveraging an access-counter set shared across MIG partitions. The second resides in software, exploiting the kernel module shared between tenants: its interrupt-handling path becomes contended when concurrent processes raise migration interrupts. The following sections detail the challenges, implementation, and evaluation of each channel.

\subsubsection{Access Counter Set-based Covert Channel}
\label{subsubsec:counter_pool_saturation}

Our reverse engineering results (\S\ref{sec:reverse}) reveal that the access counter hardware comprises 256 access counters organized into 8 sets of 32 counters each, as shown in Fig.~\ref{fig:eviction_set_size}, governed by a Least Recently Used (LRU) eviction policy. In our set-based covert channel, the sender and receiver communicate through these access counter sets. Both parties construct eviction sets: the sender accesses a set to evict any counter allocated to the receiver's process, while the receiver probes its target region to drive the counter value toward the migration threshold ($N$) and observe whether the page is subsequently migrated.

\noindent\textbf{Challenges.}
As shown in Fig.~\ref{fig:xor_masks}, counter assignment is determined by three XOR hash functions operating on physical address bits 21--38, with each 2\,MB physical memory region mapped to a single counter. This gives rise to two principal challenges.

\textit{Address opacity.} Neither the sender nor the receiver has knowledge of physical addresses~\cite{shutemov2015pagemap,khlebnikov2015pagemap,seaborn2015rowhammer}. Lacking this information, both parties are effectively blind, and assembling a known set of counters becomes a blind search. This difficulty is compounded by the migration mechanism itself: once a page is migrated from the CPU to the GPU, it is removed, preventing reliable collection of the corresponding addresses. While we employ a modified driver to reverse-engineer the counter structure, this is a controlled experimental setup and is not available to a real attacker.

\textit{Synchronization.} The sender and receiver must coordinate along two dimensions: (1)~\emph{set agreement}, ensuring both parties operate on the same counter set, and (2)~\emph{timing alignment}, ensuring transmitted and sampled bits are aligned across the two processes.

\noindent\textbf{Implementation.} We now detail our approach to enabling reliable, counter-set-based covert channel communication by overcoming the aforementioned challenges. The primary obstacle, the inability to discover exact physical addresses, cannot be resolved directly, as no architectural mechanism exposes this information to unprivileged parties. Our key observation is that the GH200 Linux kernel (\texttt{linux-nvidia-64k}) enforces a default page size of 64\,KB. This aligns perfectly with the hardware's 64\,KB migration granularity, guaranteeing coherence across all processing units in the system. Leveraging this, we look to huge pages: rather than standard 2\,MB huge pages, the GH200 platform supports 512\,MB huge pages. While a 512\,MB huge page exposes physical address bits only up to bit~29 ($512\,\text{MB} = 2^{29}$) rather than the necessary bits up to 38, we can exploit the behavior of the underlying hash functions. Specifically, these functions map each contiguous 512\,MB region onto an identical collection of 2\,MB-aligned addresses. As a result, all $256 \times 2$\,MB addresses within this region predictably map to the complete set of 256 counters. Consequently, the sender and receiver can construct their eviction sets entirely within a single 512\,MB huge page. Although this yields a functional eviction set, its absolute set ID remains opaque. Thus, the remaining challenge is to identify and synchronize corresponding sets across distinct 512\,MB huge pages.

\begin{algorithm}[t]
\caption{Receiver for Set-based Covert Channel} \label{alg:receiver_set_based}
\begin{algorithmic}[1]
\footnotesize
\State // $receiver\_pages[\,]$ is the array of pages probed each round
\State // $N$ is the threshold; $\delta_n$ is the small difference in accesses
\State // $T_{mig}$ is the migration-detection latency threshold
\State //$D$ in-between delay; $W_{det}$ migration-detection wait; $T_{idle}$ round period;  $page \gets 0,\; bit \gets 0$

\While{$bit < num\_bits$}
    \State $roundStart \gets clock();$
    \State $p \gets receiver\_pages[page];$
    \For{$k \gets 0$ \textbf{to} $N-\delta_n$}
        \State $load(p);$
    \EndFor
    \State $delay(D);$
    \For{$k \gets 0$ \textbf{to} $\delta_n$}
        \State $load(p);$
    \EndFor
    \State $delay(W_{det});$
    \State $start \gets clock();\; load(p);\; membar();\; end \gets clock();$
    \State $migrated \gets (end-start) < T_{mig};$
    \If{$migrated$}
        \State $Messages[bit] \gets \texttt{'0'}$
    \Else
        \State $Messages[bit] \gets \texttt{'1'}$
    \EndIf
    \State $bit \gets bit + 1;\; page \gets page + 1$
    \State \textbf{wait until} $clock() - roundStart \geq T_{idle};$
\EndWhile
\end{algorithmic}
\end{algorithm}

\begin{algorithm}[t]
\caption{Sender for Set-based Covert Channel} \label{alg:sender_set_based}
\begin{algorithmic}[1]
\footnotesize
\State // $d\_pages[\,]$ holds the eviction-set pages; thread $tid$ is thread ID
\State // $messages[\,]$ is the bit string to transmit; $num\_bits$ its length
\State // $max\_steps$ accesses per eviction set; $E$ number of eviction sets
\State // $T_{idle}$ bit period;
\State $kk \gets 0,\; bit \gets 0,\; setId \gets 0$
\Comment{$C$ = counters per huge page}

\While{$bit < num\_bits$}
    \State $ roundStart \gets clock();$
    \If{$messages[bit] = \texttt{'1'}$}
        \State $p \gets d\_pages[tid + C \cdot setId];$ 
            \State $kk \gets 0$
            \While{$kk < max\_steps$}
                \State $load(p);\; kk \gets kk + 1$
            \EndWhile
            \State $\mathbf{syncthreads}();$
            \State $setId \gets (setId + 1) \bmod E $
            \State $p \gets d\_pages[tid + C \cdot setId];$
            
        \State \textbf{wait until} $clock() - roundStart \geq T_{idle};$ 
    \Else  \Comment{Bit '0': stay idle for the whole period, evicting nothing}
        \State \textbf{wait until} $clock() - roundStart \geq T_{idle};\;$
    \EndIf
    \State $kk \gets 0;\; bit \gets bit + 1;\; \mathbf{syncthreads}();$

\EndWhile
\end{algorithmic}
\end{algorithm}

The algorithms for the receiver and sender of the set-based covert channel are presented in Algorithms~\ref{alg:receiver_set_based} and~\ref{alg:sender_set_based}, respectively.
The sender accesses its own eviction sets to transmit a `1' and stays idle to transmit a `0'. A key constraint is that the sender cannot repeatedly access a single set across iterations: hammering one set, i.e., 32 counters, builds up access values and eventually migrates the sender's own pages. While LRU should in principle prevent this, \papertitle{} issues accesses to each distractor page in parallel, and because GPU threads execute non-deterministically, build-up can still occur. To avoid this, the \papertitle{} sender accesses different eviction sets in an alternating fashion.
The receiver attempts to trigger a migration by completing $N$ accesses to its own pages in two steps. It first issues $N-\delta$ accesses, then waits for a duration $D$ to allow the sender to signal, and finally completes the remaining $\delta$ accesses. After all accesses finish, the receiver waits for a migration: if a migration occurs, the transmitted bit is `0'; otherwise, it is `1'. 

\textbf{Set Synchronization.}
We describe three methods to establish agreement.

\emph{Method 1: Allocation ordering.} One process first creates a pool of huge pages. In \papertitle{}, a process can only request 512\,MB of physically contiguous memory, which the buddy allocator satisfies from an available page. If a huge page is allocated and then released back to the buddy allocator intact (preserving its VA-to-PA mapping), it can be re-allocated by the next requesting process. Thus a process that allocates $M$ huge pages and releases them in a known order allows a subsequent process, after the first terminates, to re-allocate the same pages in reverse order. The first process therefore sorts out the eviction sets, and the sender and receiver reuse these pages. This method assumes no other process allocates a huge page in between.

\emph{Method 2: Set-based synchronization.} The receiver collects its pages, and the sender, by listening to the receiver, determines the exact set the receiver is using. This is analogous to the cache-set synchronization procedure.

\emph{Method 3: Exhaustive coverage.} Since there are only 8 sets, and the receiver accesses only a single page while trying to migrate it, the sender can touch all eviction pages across parallel thread blocks. This works because the receiver spends more time waiting for migration, while the sender merely accesses pages in parallel, taking less time and triggering no migration of its own.

\begin{figure}
    \centering
    \includegraphics[width=1\linewidth]{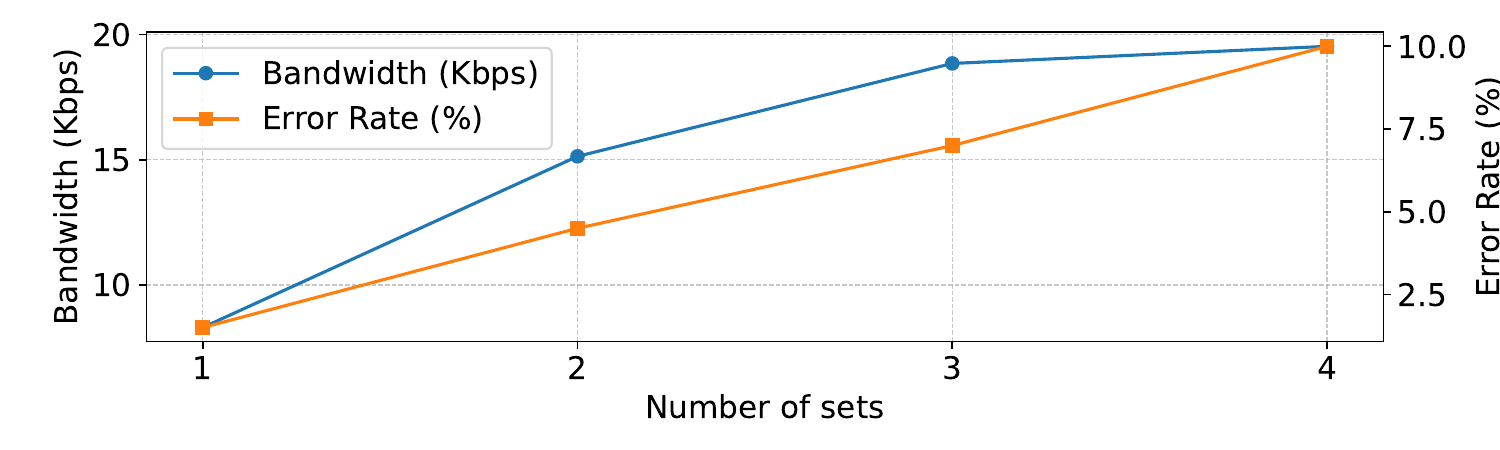}
    \caption{Eviction-based covert channel evaluation.}
    \label{fig:covert_result_parallel_sets}
\end{figure}

\textbf{Evaluation and Results.} The error rate is computed using the Levenshtein edit distance~\cite{miller2009levenshtein}. The sender transmits a message of 10{,}000 randomly generated bits, and we compare the received bit sequence against the ground-truth message to obtain the error rate. The set-based covert channel can also be extended across parallel sets. Figure~\ref{fig:covert_result_parallel_sets} reports the resulting bandwidth and error rate. The bandwidth nearly doubles when moving from a single set (8{,}500~bps) to two sets (15{,}500~bps), but it does not scale linearly with the number of sets. This sublinear scaling arises because a receiver probing multiple pages introduces contention on the driver's interrupt-handling and migration pipeline. As the number of migration requests grows, each probe must wait longer due to the increased contention, which in turn raises the error rate.

\subsubsection{Migration Delay Variation}
\label{subsubsec:migration_delay_variation}
This covert-channel primitive leverages the fact that the hardware-assisted
page migration delay can be varied based on the concurrent GPU context. The delay
arises when a concurrent process delays the NVIDIA driver from serving a
migration request issued by a process that has made enough accesses to a page
to trigger promotion. This heightened delay can originate from memory allocation
APIs, from the UVM driver servicing faults, or from OS-related stalls, as
explained in Section~\ref{sec:reverse}.

\textbf{Implementation.}
This covert channel follows the set-based
covert channel (Algorithms~\ref{alg:receiver_set_based}
and~\ref{alg:sender_set_based}); indeed, the receiver is identical to the
set-based case. The sender is modified to also migrate pages: it raises its
access count to $N$ and does not wait for the migration to complete. To observe
the channel, the receiver simply measures the migration delay, the interval
from when migration is requested until it actually occurs. The sender encodes
bits by creating contention: it delays the receiver's migration by performing
contending concurrent accesses. 

\textbf{Evaluation and Results. }The evaluation follows the same methodology as the eviction-set-based covert channel. This channel achieves approximately 8{,}100 bps with a 4\% error rate. Its main drawback is that it relies purely on software-level contention, which makes it unstable across runs, primarily because the contention level is difficult to control. We measure the \textit{sensitivity} of this covert channel, defined as the fraction of runs in which the message is transmitted successfully. This channel succeeds in roughly half of all runs. Thus, while it is convenient to implement, it suffers from inherent instability.

\subsection{Side Channel Attacks}
\label{subsec:side}

This section outlines side channel leakages across different proof-of-concept attacks; fingerprinting CUDA applications~\cite{nvidia_cuda_samples},  SPEC2017 CPU benchmarks~\cite{spec2017}, Large Language Models~\cite{wolf2020transformers, huggingface_hub}, and LLM input-output token length estimation leveraging the migration delay leakage. We also exploit counter-set leakage as well, showing that different Polybench kernels~\cite{pouchet2012polybench} from UVMBench~\cite{gu2020uvmbench, osu_uvm_benchmark} can be identified by the leakage if they are using system-allocated memory. \papertitle{} shows a high success rate for these attacks.

\begin{figure*}[tbh]
    \centering
    \subcaptionbox{503.bwaves\_r~\cite{spec2017,bucek2018spec}\label{fig:503}}{%
        \includegraphics[width=0.33\textwidth]{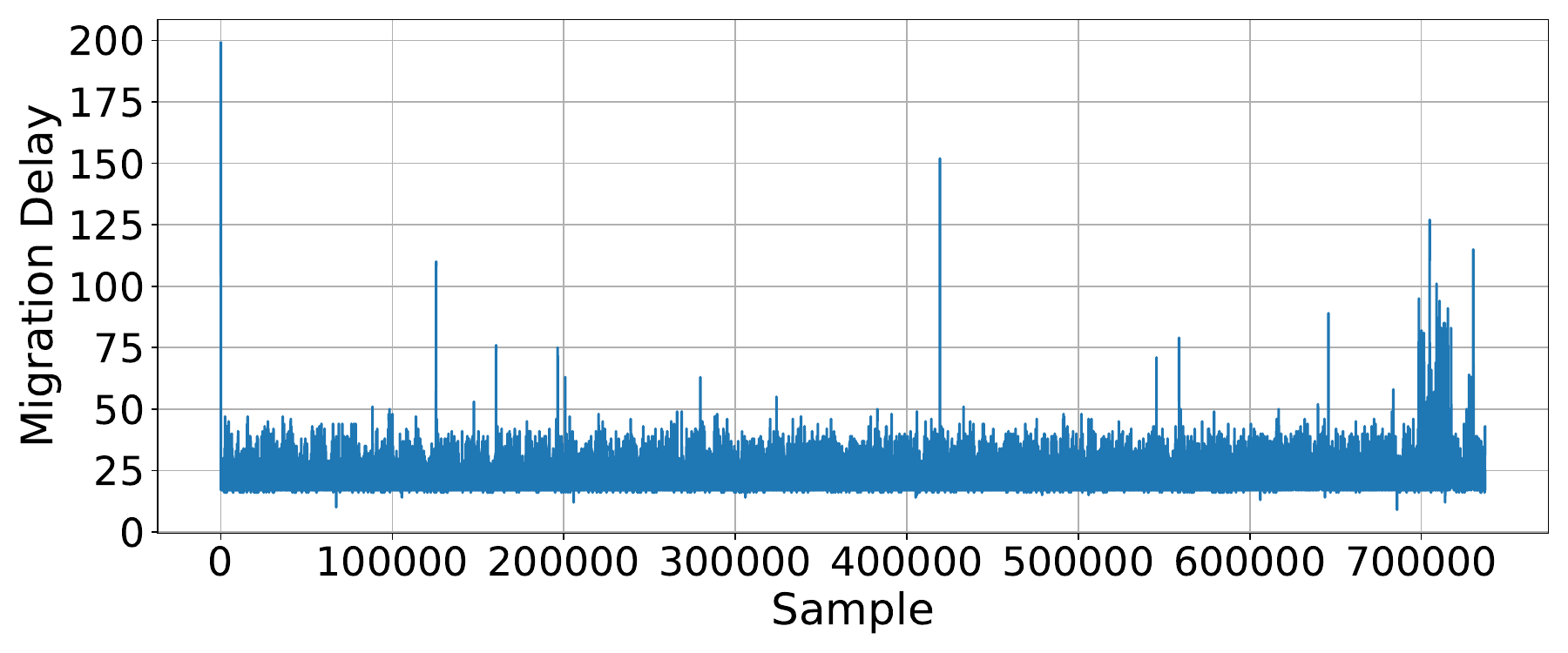}}%
    \hfill
    \subcaptionbox{549.fotonik3d\_r\label{fig:549}}{%
        \includegraphics[width=0.33\textwidth]{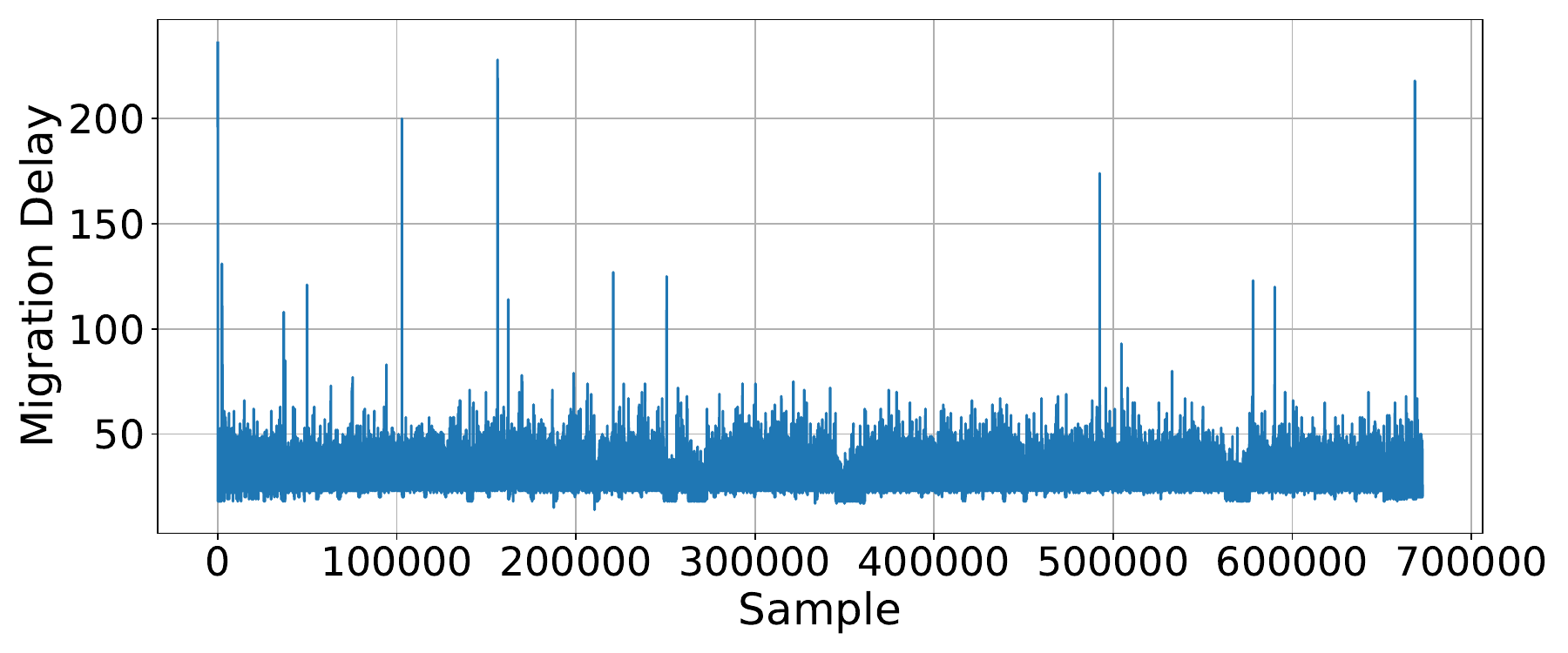}}%
    \hfill
    \subcaptionbox{603.bwaves\_s\label{fig:srad}}{%
        \includegraphics[width=0.33\textwidth]{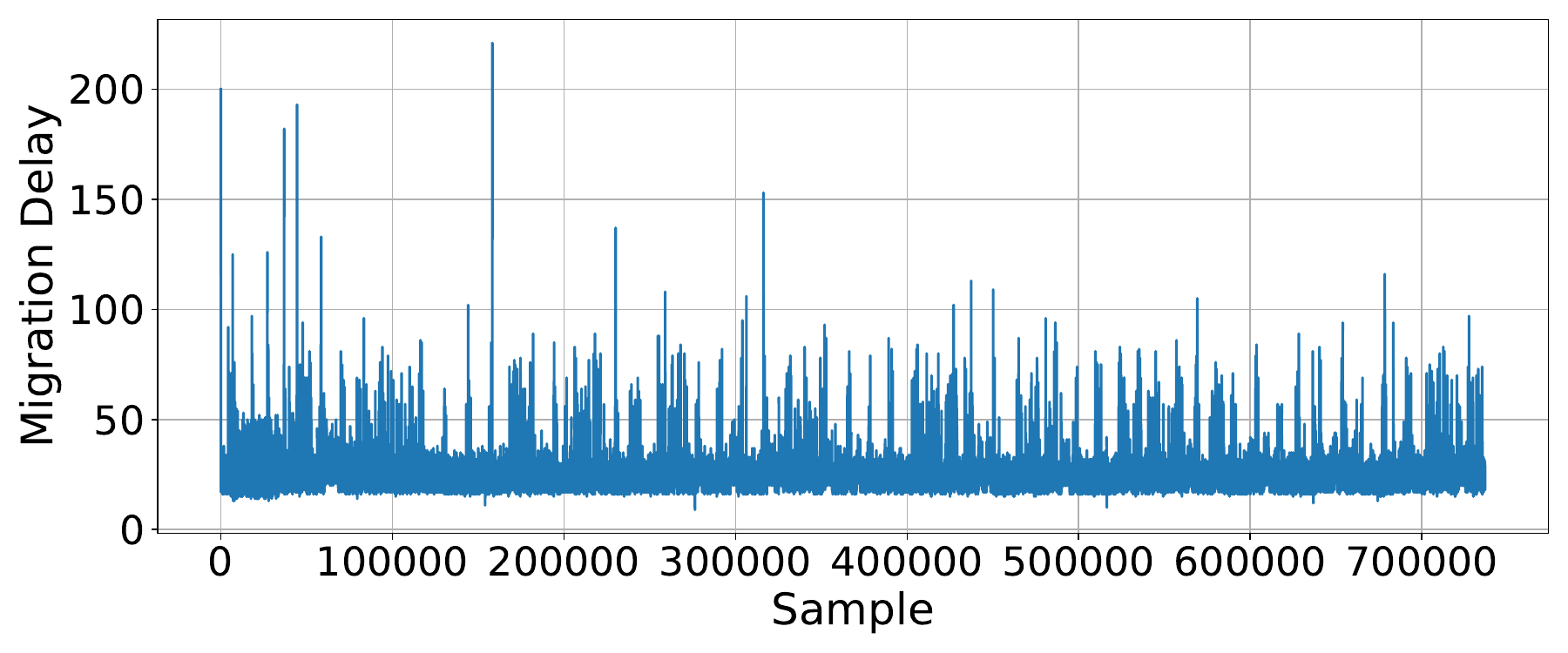}}
    \\[0.5em]
    \subcaptionbox{Bart Large~\cite{lewis2020bart}\label{fig:blender1}}{%
        \includegraphics[width=0.33\textwidth]{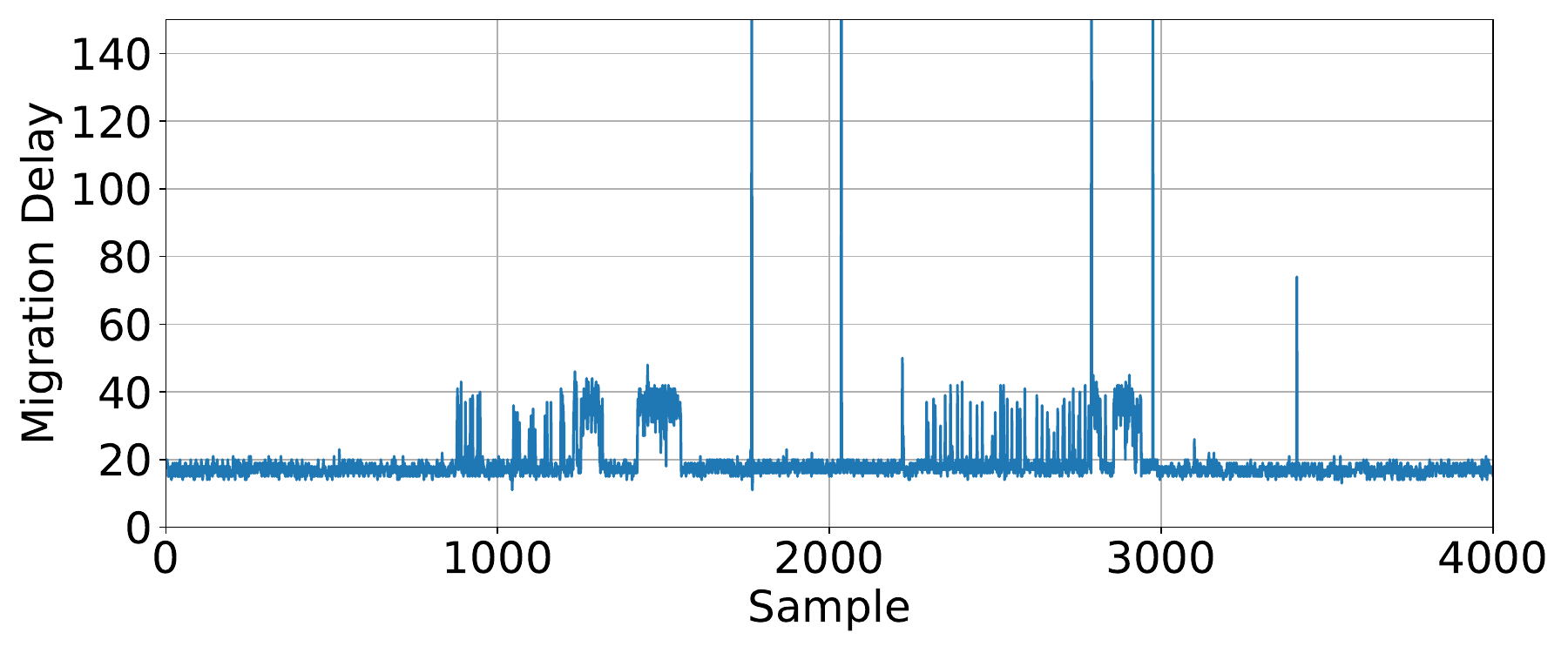}}%
    \hfill
    \subcaptionbox{Llama~\cite{touvron2023llama}\label{fig:blender2}}{%
        \includegraphics[width=0.33\textwidth]{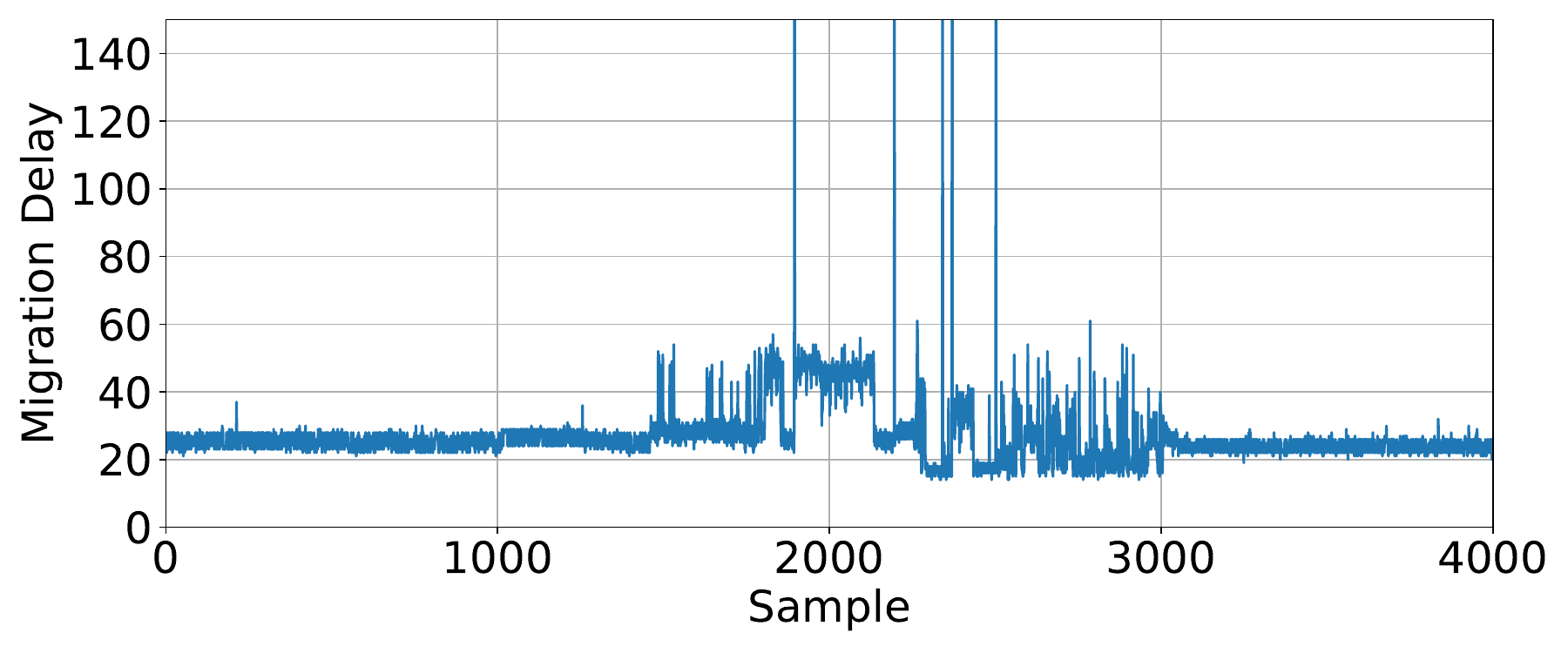}}%
    \hfill
    \subcaptionbox{Qwen3~\cite{yang2025qwen3}\label{fig:qwen3}}{%
        \includegraphics[width=0.33\textwidth]{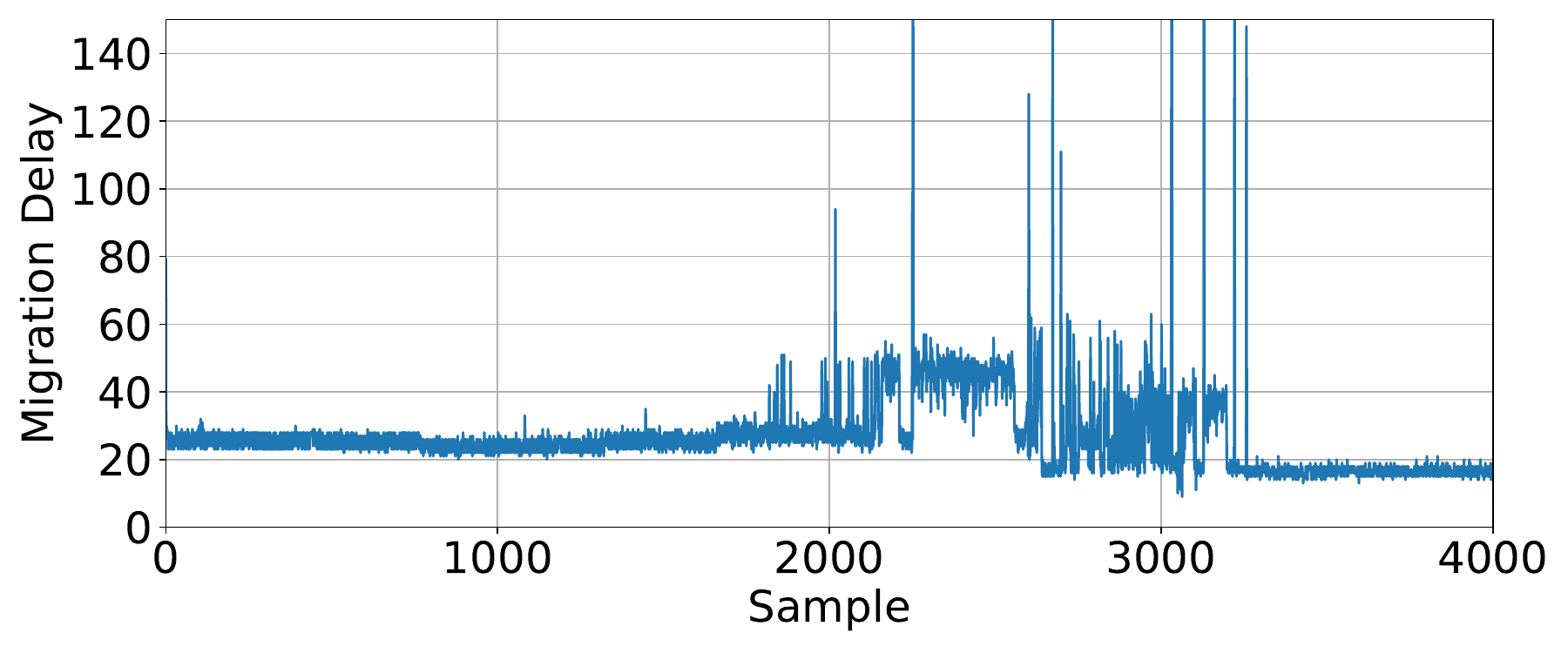}}
    \caption{Migration delays observed by the spy while the victim is running CPU workloads or LLM inference tasks.}
    \label{fig:plots_spec_llm}
\end{figure*}

\subsubsection{General Methodology}
\label{subsubsec:gen_method}
In an offline phase, the spy collects data samples for the applications under consideration and trains machine learning models to classify them.  We evaluate models including random forest~\cite{breiman2001random}, histogram-based gradient boosting~\cite{ke2017lightgbm}, XGBoost~\cite{chen2016xgboost}, LightGBM~\cite{ke2017lightgbm}, a soft voting ensemble~\cite{scikit-learn}, and a convolutional neural network (CNN)~\cite{lecun1998gradient} using the feature engineering (Table~\ref{tab:features}). In the online phase, it collects the side channel leakage and uses the trained models to identify which victim application is running. 

\textbf{Threat Model.} The attacker and victim reside on the same platform but are isolated as described in Section~\ref{subsec:threat_model}: each is assigned a distinct MIG instance. The spy migrates system-allocated pages while the victim executes its own workloads. We assume the spy holds only user-level privileges and can launch CUDA kernels within its own instance, including the use of all user-accessible registers such as \texttt{CLOCK}. For CPU workload fingerprinting, the victim need not be using the GPU at all.

\subsubsection{Migration Delay Leakage: CUDA Applications}
\label{subsubsec:cuda_app}

\begin{table}[htbp]
\centering
\caption{CUDA Applications: five-fold cross-validation performance
over 25 classes. Values are mean (standard deviation) across folds, in percent.}
\label{tab:cuda_results}
\begin{tabular}{lccc}
\hline
 & \textbf{F1 (\%)} & \textbf{Prec (\%)} & \textbf{Rec (\%)} \\
 & $\mu\,(\sigma)$ & $\mu\,(\sigma)$ & $\mu\,(\sigma)$ \\
\hline
Random Forest & 90.1 (0.5) & 90.7 (0.7) & 90.4 (0.4) \\
XGBoost       & 93.6 (0.6) & 93.9 (0.6) & 93.6 (0.6) \\
\hline
\end{tabular}
\end{table}

Different CUDA applications communicate with the driver in distinct ways. Characterizing these differences across applications (drawn from the NVIDIA CUDA Samples~\cite{nvidia_cuda_samples}) enables an attacker to distinguish GPU workloads and target one that may contain a secret.

\textbf{Mechanism.} As the spy measures migration delay, this measurement yields a unique fingerprint of how the victim interacts with the GPU driver, such as through memory-management APIs, kernel launches, and teardown.

\textbf{Methodology and Results.} We collect 100 traces for each of the 25 CUDA applications (Table~\ref{tab:cuda_apps}), yielding 2500 traces in total. Each collection runs for around 8 seconds and gathers 131{,}000 samples. Table~\ref{tab:cuda_results} reports the fingerprinting results, achieving nearly 94\% accuracy in identifying which CUDA application the victim is running.

\subsubsection{Migration Delay Leakage: SPEC2017 Benchmarks}
\label{subsubsec:spec_app}

Unlike leakage across GPU workloads, inferring information about CPU workloads poses a broader threat for two reasons. First, every co-located process leaves a CPU footprint, whereas only a subset of workloads exercises the GPU; CPU-side fingerprinting therefore applies to essentially any victim sharing the platform. Second, the CPU execution profile of a workload, i.e. its pattern of OS interactions such as system calls, page faults, scheduling behavior, and other kernel-level activity, is closely tied to the program's identity and phase structure, so distinguishing among workloads can reveal what a co-tenant is running and, by extension, information about its function or input. By combining the spy's ability to infer both CPU and GPU activity, the adversary can correlate the two signals to refine its inferences.

\textbf{Mechanism.} During the service stage, the driver must read the process's memory map, which requires the \texttt{mmap\_lock}. That lock prioritizes writers. So if any CPU thread in the same process is modifying memory (via \texttt{malloc}/\texttt{free}, \texttt{mmap}, \texttt{munmap}, \texttt{mprotect}, etc.), its write request blocks new readers, including the migration, until it finishes. Because the migration stages run strictly one after another, this delay adds to the total time between a remote access and the data actually moving. Hence, heavy CPU memory activity (e.g. a large \texttt{munmap} tearing down many memory regions) can stall the GPU's migration work.

\textbf{Methodology and Results. }We collect 100 traces per benchmark (Table~\ref{tab:spec2017}), yielding 2500 traces in total. For each trace, the spy collects around 737{,}000 samples over 45 seconds. Table~\ref{tab:spec_results} reports the performance of the attack across 5 different techniques,
achieving up to 89\% accuracy (XGBoost, LightGBM, and Soft-voting Ensemble) in identifying the victim's CPU workload. Figure~\ref{fig:plots_spec_llm} shows time-series plots for 3 SPEC benchmarks (503.bwaves\_r, 549.fotonik3d\_r, and 603.bwaves\_s). Overall, the plots exhibit readily distinguishable patterns.

\begin{table}[htbp]
\centering
\caption{SPEC2017:  Five-fold cross-validation (mean, and std. dev macro-averaged across folds).}
\label{tab:spec_results}
\begin{tabular}{lccc}
\hline
 & \textbf{F1  (\%)} & \textbf{Prec  (\%)} & \textbf{Rec  (\%)} \\
 & $\mu\,(\sigma)$ & $\mu\,(\sigma)$ & $\mu\,(\sigma)$ \\
\hline
Random Forest & 84.8 (1.9) & 85.4 (1.7) & 84.8 (1.8) \\
HistGradientBoosting & 86.4 (0.7) & 86.8 (0.8) & 86.5 (0.6) \\
XGBoost & 88.6 (0.9) & 89.1 (0.9) & 88.7 (0.8) \\
LightGBM & 89.0 (1.0) & 89.6 (1.1) & 89.1 (1.0) \\
Soft-voting Ensemble & 88.6 (1.2) & 89.0 (1.3) & 88.6 (1.2) \\
\hline
\end{tabular}
\end{table}

\subsubsection{Large Language Model Architecture Fingerprint}
\label{subsubsec:llm_arch}

Our attack aims to identify the specific Large Language Model (LLM) underlying a target system. Knowing which model is deployed carries significant consequences for both security and privacy. First, because LLMs are often fine-tuned for particular domains, pinpointing the model in use can indirectly expose the system's intended function. Second, since every model carries its own characteristic set of weaknesses, limitations, and failure modes, accurate identification enables an attacker to refine subsequent exploits, for instance, by crafting adversarial inputs tailored to that model. Finally, in adaptive systems that route queries to different models depending on the input~\cite{routerbench}, learning which LLM handled a given request can reveal sensitive inference-level information about the user's prompt.

\textbf{Methodology and Results.} We collect 100 traces per model across 20 LLMs listed in Table~\ref{tab:llms}, yielding 2000 traces in total. For each trace, the spy collects around 120{,}000 samples over 48 seconds. Although the sampling rate is lower here, Table~\ref{tab:llm_results} reports the performance of the attack across 5 different techniques, achieving up to 94\% accuracy (LightGBM) in identifying the victim's workload. Figure~\ref{fig:plots_spec_llm} shows time-series plots for 3 LLMs (Bart Large, Llama, and Qwen3); the LLM plots are zoomed into the region that reveals the CUDA context launch patterns. In the next attack, we show how an attacker can leverage the full trace for token-size estimation. 

\begin{table}[htbp]
\centering
\caption{LLM: Five-fold cross-validation (mean, and std. dev macro-averaged across folds).} 
\label{tab:llm_results}
\begin{tabular}{lccc}
\hline
 & \textbf{F1 (\%)} & \textbf{Prec (\%)}  & \textbf{Rec (\%)} \\
 & $\mu\,(\sigma)$ & $\mu\,(\sigma)$ & $\mu\,(\sigma)$ \\
\hline
Random Forest & 91.9 (1.0) & 92.3 (1.0) & 92.0 (1.1) \\
HistGradientBoosting & 91.9 (0.9) & 92.4 (0.8) & 92.0 (0.9) \\
XGBoost & 92.6 (1.3) & 92.9 (1.2) & 92.6 (1.3) \\
LightGBM & 93.7 (1.1) & 94.0 (1.0) & 93.7 (1.1) \\
Soft-voting Ensemble & 93.3 (1.0) & 93.6 (0.9) & 93.3 (0.9) \\
\hline
\end{tabular}
\end{table}

\begin{figure}[htbp]
    \centering
    \includegraphics[width=1\linewidth]{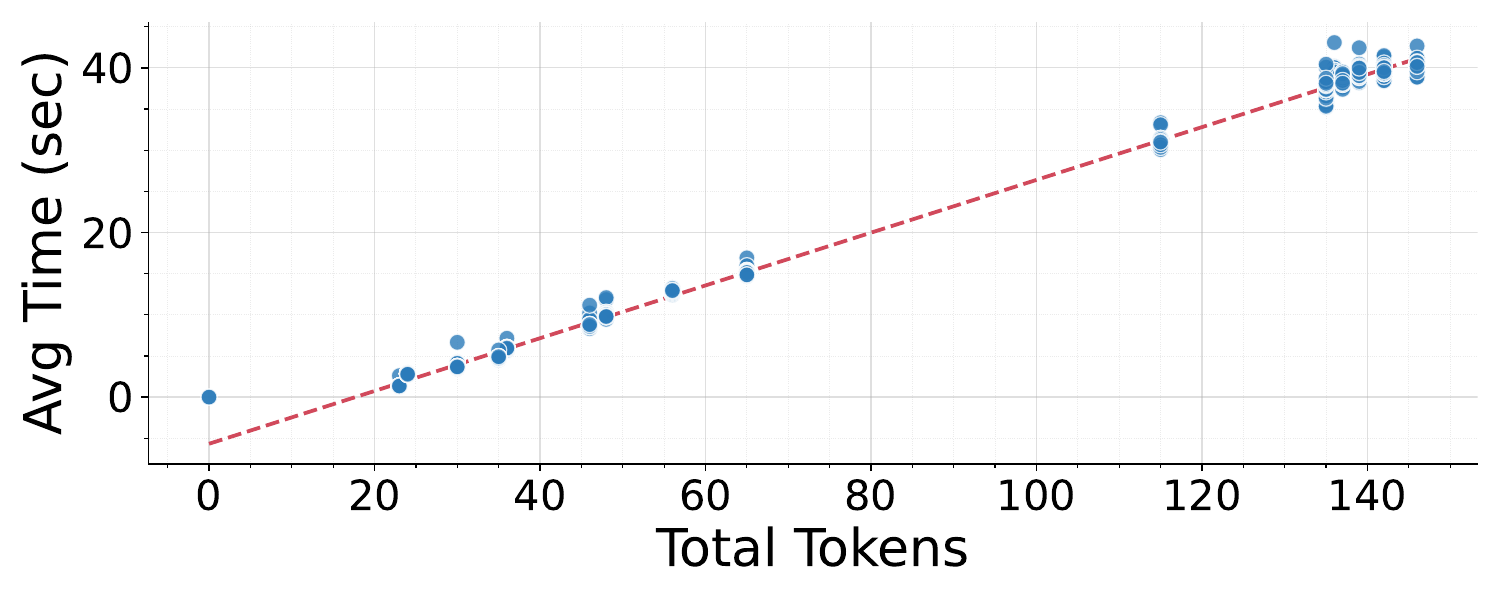}
    \caption{Response time vs. processed token length.}
    \label{fig:token_length_est}
\end{figure}

\subsubsection{Generalization Across Cloud Providers}
To evaluate whether model fingerprints transfer across different runtime environments, we train our classifier on side-channel traces collected from an NVIDIA GH200 instance on CloudLab~\cite{Duplyakin+:ATC19} (GPU Driver: 590.48.01; CUDA toolkit: 12.2) and evaluate it on traces collected from a separate GH200 instance hosted by Lambda (GPU Driver: 580.65.06; CUDA toolkit: 13.0) across the same 20 models. 
For each trace, we extract 65 features capturing the model-loading burst, including its onset timing along a cumulative-delay axis, the post-onset activity profile, and the number of active memory regions. 
Features are standardized independently per machine to account for baseline platform discrepancies. 
We train an Extra-Trees classifier (500 estimators) exclusively on the CloudLab dataset (80 traces per model) and evaluate it on the Lambda dataset (20 traces per model). 
The classifier achieves 72\% cross-provider accuracy, demonstrating that the extracted side-channel signatures capture underlying, machine-independent structural characteristics of the models rather than platform-specific noise.

\subsubsection{Input-Output Prompt Length Estimation} LLM inference proceeds in two phases. In the prefill phase, the model processes the entire input prompt in a single forward pass, initializing the key-value (KV) cache. In the decoding phase, it generates output tokens one at a time, each reusing the cached context, until a stopping criterion is met. The phases have different bottlenecks: prefill is compute-bound, performing large parallel matrix multiplications, while decoding is memory-bandwidth-bound, repeatedly reading the weights and growing KV cache to emit one token per step. As a result, prefill runs as a few long kernels, whereas decoding issues many short ones and dominates total latency, since each additional output token adds another serial step. Figure~\ref{fig:token_length_est} shows that the execution time is linearly dependent on the number of input-output tokens, based on 20 prompts processed by Google's Gemma 3 1B model~\cite{gemma_2025}. This stems mainly from the decode phase's increasing execution time due to its sequential processing. 

This observation enables an attacker who can identify the start and end of these phases from the side channel leakage to estimate the duration of inference. Given that the attacker has found out the LLM leveraged in the serving system, the attacker now moves forward to estimate the length of the input/output prompt processed in the inference pass. We assume the processing batch size is 1; hence, the time taken by the victim's prompt inference is related to the number of output tokens.

\begin{figure}[htbp]
    \centering
    \begin{tikzpicture}
        \node[anchor=south west, inner sep=0] (image) at (0,0)
            {\includegraphics[width=1\linewidth]{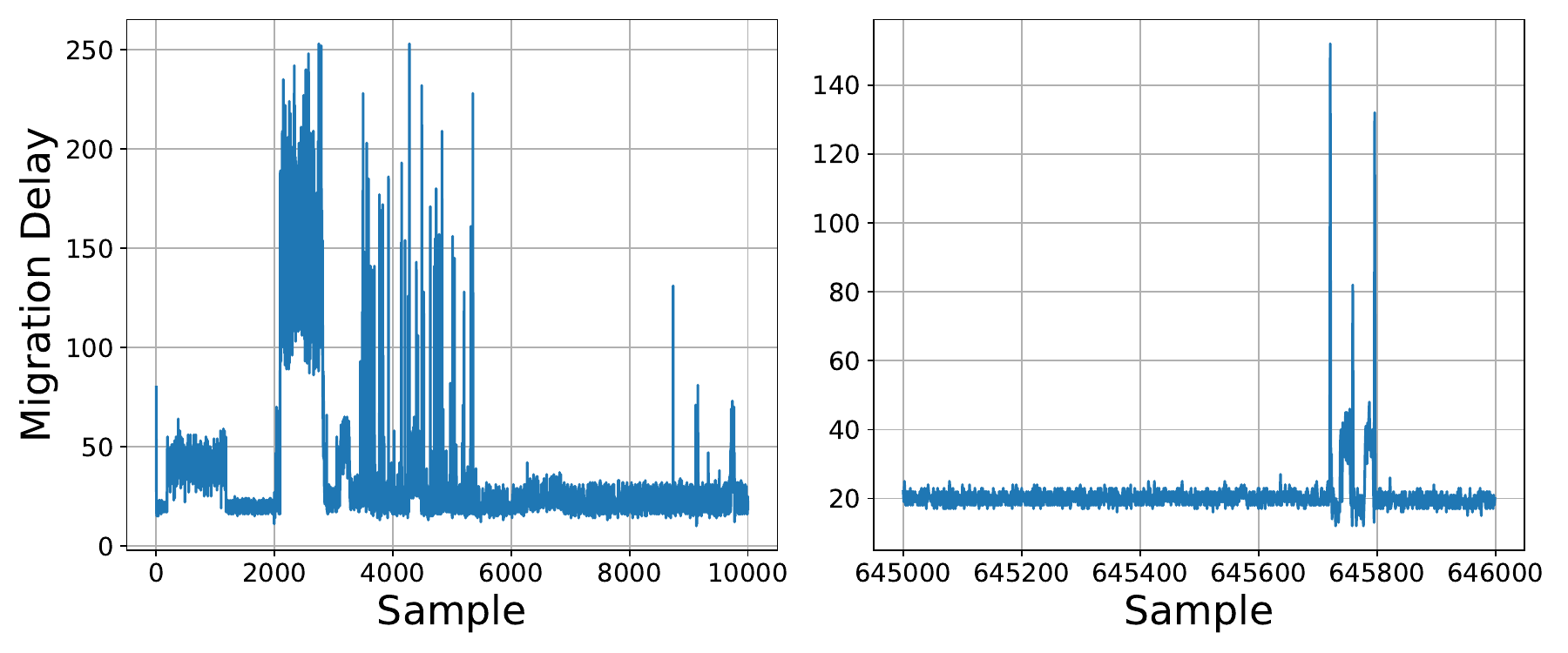}};
        
        \begin{scope}[x={(image.south east)},y={(image.north west)}]
        
            \draw[red, dashed, thick] (0.1,0.18) -- (0.1,0.78);
            \draw[red, dashed, thick] (0.32,0.18) -- (0.32,0.78);
            
            \draw[->, red, thick] (0.40,0.9) -- (0.225,0.80);
            \node[red, font=\small] at (0.40,0.95) {CUDA context launch};
            
            \draw[->, blue, thick] (0.55,0.55) -- (0.46,0.30)
                node[blue, font=\small]at (0.55,0.60) {prefill finish};
            
            \draw[red, dashed, thick] (0.84,0.20) -- (0.84,0.80);
            \draw[red, dashed, thick] (0.89,0.20) -- (0.89,0.80);
            
            \draw[->, red, thick] (0.93,0.90) -- (0.86,0.78)
                node[red, font=\small, anchor=east] at (1,0.93) {decode finish};
        
        \end{scope}
    \end{tikzpicture}
    \caption{Migration delay for Google Gemma 3 1B~\cite{gemma_2025} for input prompt ``Complete: The quick brown fox...''.}
    \label{fig:prefill_decode}
\end{figure}

\textbf{Methodology and Results. } Similar to the other attacks, this PoC has an \emph{offline} phase for data collection (around 737{,}000 samples over 47 seconds) and characterizing the system's response behavior, and an \emph{online} phase in which the attacker estimates the processed token length. We increase the sampling rate here to accurately capture the tear-down signal. Figure~\ref{fig:prefill_decode} shows Gemma~3's behavior (similar across other LLMs) in response to the prompt \texttt{``Complete: The quick brown fox\ldots''}. From the side-channel leakage, we distinguish two phases, \emph{prefill} and \emph{decode}: we first observe the CUDA context launch (GPU process start), then a prefill region up to roughly sample~10{,}000, followed by a characteristic tear-down pattern marking decode completion. Mapping this against the average-time-versus-total-token analysis, the onset of the tear-down pattern tracks inference time: longer inferences push the tear-down onset later in the trace, while shorter inferences make it appear earlier.
We collect migration delay side channel traces for 20 prompts on the Gemma~3 1B model. In the offline phase, we learn the system's response to varying prompt lengths in order to (1)~measure the average time taken per prompt length, and (2)~characterize the tear-down pattern in the side-channel signal. In the online phase, the attacker uses two components: a \emph{pattern finder}, which locates the sample index where the tear-down pattern begins in the trace, and a \emph{linear regression model}, which maps that index to an estimate of the number of processed tokens using the relationship learned offline. We evaluate on a held-out 20\% test split, stratified by class. We fit a linear estimator on the training streams, mapping each detected tear-down onset position to its token count, and measure performance on the held-out streams using $R^2$ (coefficient of determination) and mean absolute error. The method achieves $R^2 = 0.92$ with an MAE of 4.6 tokens, recovering token counts from the raw signal alone.

\subsubsection{Counter-Set Leakage: CUDA Kernel Fingerprinting}
\begin{figure}[t]
    \centering
    \begin{subfigure}{0.5\textwidth}
        \centering
        \includegraphics[width=\linewidth]{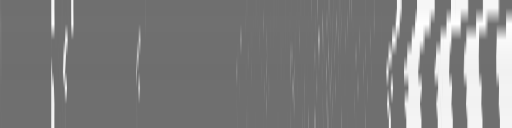}
        \caption{FDTD-2D}
        \label{fig:fdtd}
    \end{subfigure}

    \vspace{0.5em}

    \begin{subfigure}{0.5\textwidth}
        \centering
        \includegraphics[width=\linewidth]{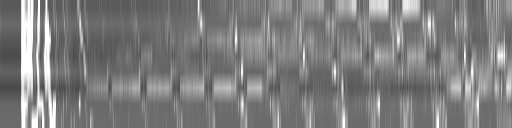}
        \caption{SYR2K}
        \label{fig:syr2k}
    \end{subfigure}

    \vspace{0.5em}

    \begin{subfigure}{0.5\textwidth}
        \centering
        \includegraphics[width=\linewidth]{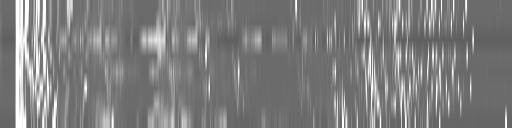}
        \caption{SYRK}
        \label{fig:syrk}
    \end{subfigure}

    \caption{Polybench heatmaps.}
    \label{fig:polybench_heatmaps}
\end{figure}

For this attack, we use Polybench~\cite{pouchet2012polybench} kernels drawn from the UVMBench~\cite{gu2020uvmbench,osu_uvm_benchmark} benchmark suite. To ensure that the victim applications exercise the counters, we modify their memory allocation APIs to use \texttt{malloc}. We probe across eight sets to determine whether the spy's counter has been evicted, which we infer by checking whether the page under access is successfully migrated. To suppress noise introduced by migration delay, we increase the spy's migration wait time; this reduces the effective sampling rate to approximately 2{,}000 samples per set per second. Each observation is recorded as `0' when migration is possible and `1' when it is not. After collecting data for the ten benchmarks listed in Table~\ref{tab:polybench} over a 50-second window (100K samples per set), we apply several model-based analyses to evaluate classification accuracy, including a CNN trained on the corresponding heatmaps (Figure~\ref{fig:polybench_heatmaps}). Using a soft-voting ensemble, we achieve close to 95\% success in identifying the kernel executed by the victim, as reported in Table~\ref{tab:polybench_results}.

\begin{table}[htbp]
\centering
\caption{Eviction-Set based attack: Five-fold cross-validation performance (macro-averaged). Values are the mean (standard deviation) across folds, in percent.}
\label{tab:polybench_results}
\begin{tabular}{lccc}
\hline
 & \textbf{F1 (\%)} & \textbf{Prec (\%)} & \textbf{Rec (\%)} \\
 & $\mu\,(\sigma)$ & $\mu\,(\sigma)$ & $\mu\,(\sigma)$ \\
\hline
Random Forest & 93.6 (1.9) & 93.8 (1.8) & 93.6 (1.9) \\
HistGradientBoosting & 92.4 (1.6) & 92.8 (1.6) & 92.4 (1.5) \\
XGBoost & 92.3 (1.8) & 92.7 (1.8) & 92.4 (1.8) \\
LightGBM & 92.8 (2.3) & 93.0 (2.2) & 92.8 (2.2) \\
Soft-voting Ensemble & 94.5 (0.0) & 94.5 (0.0) & 94.5 (0.0) \\
Heatmap CNN & 85.5 (5.7) & 86.7 (5.5) & 85.9 (5.5) \\
\hline
\end{tabular}
\end{table}

\subsubsection{Future Finer-Grained Attacks} While PAGE404 primarily demonstrates coarse-grained fingerprinting, the vulnerability can be extended to fine-grained access metering due to an architectural granularity mismatch and persistent tracking state. Because page migration counters are tracked per 2\,MB region while physical pages are allocated at 64\,KB boundaries, distinct pages within the same aligned 2\,MB range aggregate accesses to a single shared counter that persists across execution boundaries. An adversary can exploit this behavior by accessing a co-located 64\,KB page to prime the shared counter close to the migration threshold, allowing the victim to execute, and subsequently probing their own page in small increments until migration is observed. By measuring the access delta required to trigger migration, the attacker transforms a binary access-detection channel into a quantitative side channel that can infer the victim's precise access count within a 5\% margin of error.

\section{Potential Mitigations}
\label{sec:mitigation}

The root of the vulnerability is the shared counter hardware structure, and the shared access to the driver code.  This sharing enables contention and causes leakage through observable delays.  One approach for mitigation is to carry the MIG philosophy to ensure isolation even through these shared components in software and hardware.   With respect to hardware isolation, this requires partitioning the counters into groups of counters dedicated to each MIG partition rather than sharing the counters across them.  To close the contention through the driver, performance isolation of the driver, again perhaps by MIG partition, can mitigate the observable delays.  The vGPU abstraction~\cite{vGPU} provides driver isolation on MIG-backed versions, but NVIDIA does not yet support ARM platforms. Moreover, since the guest drivers in each instance are multiplexed by the host vGPU manager, the leakage may shift to observable delays through the vGPU manager; whether such delays are observable requires further confirmation through experimentation.
\section{Related Work}
\label{sec:related}

\textbf{Intra-GPU leakage.}
A large body of research targets \textit{intra-GPU} leakage, exploiting shared
microarchitectural resources such as caches~\cite{naghibijouybari2017constructing,
naghibijouybari2018rendered}, the memory bus~\cite{wei2020leaky}, functional units and
on-chip networks~\cite{ahn2021network, wan2022meshup}, TLB state~\cite{zhang2024invalidate+},
and DRAM-level effects~\cite{frigo2018grand, paccagnella2021lord} to leak information between
co-located GPU processes. On the defense side, GPU performance counters have been used to
detect and mitigate contention-based channels~\cite{xu2019gpuguard}
and to validate kernel execution integrity~\cite{almusaddar2025shadowscope}. However, both defenses rely on GPU performance counters, which MIG
partitions per instance, making them unlikely to detect an attacker
residing in a separate MIG instance. Extending such defenses to
operate at the host level and monitor activity across MIG instances could enable detection.

\textbf{Cross-domain (CPU--GPU) leakage.}
A complementary line of work examines \textit{cross-domain} leakage between the CPU and GPU,
exploiting resources shared across the heterogeneous boundary such as the PCIe bus and shared
memory subsystems~\cite{dutta2021leaky, side2022lockeddown}.

\textbf{Multi-GPU interconnect leakage.}
Attacks have also been extended to \textit{multi-GPU} settings, where distributed workloads
across multiple devices create observable contention on inter-GPU
interconnects~\cite{dutta2023spy}. Zhang et al.~\cite{zhang2024beyond} showed that NVLink
interconnect contention can be used to infer inter-device communication patterns across GPUs.
However, this approach targets aggregate communication behavior and relies on bandwidth
saturation --- a condition that is increasingly difficult to achieve on modern high-bandwidth
links such as the GH200's 900\,GB/s NVLink-C2C~\cite{zhang2025nvbleed}.

\textbf{Rowhammer attacks on GPUs.} Recent work has shown that Rowhammer is a practical threat to GPU memories.
GPUHammer~\cite{lin2025gpuhammer} demonstrated the first Rowhammer attacks on
GDDR memory in GPUs and showed that the resulting bit flips can degrade the
accuracy of machine learning models. Follow-up work turned these bit flips into
stronger exploits: GPUBreach~\cite{lin2026gpubreach} and
GeForge~\cite{wan2026geforge} corrupt GPU page tables to escalate privileges,
and GDDRHammer~\cite{hu2026gddrhammer} mounts cross-component attacks
originating from the GPU. Most recently, GPUThor~\cite{lin2026gputhor} uses
non-uniform hammering patterns to induce exploitable bit flips even on
ECC-protected GPUs.

\textbf{MIG isolation.}
Prior work has examined the isolation boundary between MIG instances, which NVIDIA advertises as complete and strongly hardware-enforced~\cite{nvidia_mig_userguide,nvidia2020ampere}. Zhang et al.~\cite{zhang2023t} show that the last-level TLB is shared, while recent work by Gu et al.~\cite{gubehind} shows that some leakage remains across cache partitions in the presence of memory-barrier operations.  Our work demonstrates a novel form of leakage that also compromises MIG isolation for this emerging class of tightly integrated architectures.  It is important to identify and characterize this mode of leakage to both inform interventions and guide the design of future tightly integrated systems. 
Miao et al.~\cite{miao2024veiled} extend MIG-bypassing channels into the GPU ``uncore'' (DRAM, host--GPU and inter-GPU links), without requiring root privileges. More recently, MIGraine~\cite{dimigraine} demonstrates page-fault–based leakage across isolated MIG instances, even when the instances are provisioned as separate vGPUs.

\section{Conclusion}
\label{sec:conclusion}
In this paper, we present \papertitle{}, an analysis of the emerging Grace Hopper heterogeneous computing platform and its hardware-backed page migration scheme. We reveal that the access-counter hardware structure is shared across MIG instances, violating MIG's security guarantees, and we further identify a migration-delay leakage arising from the driver's interrupt-handling pipeline. Building on these findings, \papertitle{} demonstrates a range of proof-of-concept covert- and side-channel attacks. The eviction-based covert channel achieves 8500\,bps at a 1.5\% error rate, rising to 15500\,bps at roughly 5\% error with two eviction sets, while the eviction-based side channel distinguishes among 10 CUDA kernels with 94\% accuracy when they use system-allocated memory. The migration-delay leakage enables a covert channel of 8100\,bps at a 4\% error rate and, across several side-channel attacks, distinguishes CPU, CUDA, and LLM workloads with around or above 90\% accuracy, estimating an LLM's processed token length with a mean absolute error of 4.6 tokens on average. These results show that \papertitle{} attains a high attack success rate and underscore the need for more secure architectures in multi-tenant environments.

\ifCLASSOPTIONcompsoc
  \section*{Acknowledgments}
\else
  \section*{Acknowledgment}
\fi

We sincerely thank our shepherd and anonymous reviewers for their
valuable feedback. This work was partially supported by the National Science Foundation under Grants SSF-2512416 and CPIS-2448156, and in part by the Commonwealth Cyber Initiative, an investment in the advancement of cyber R\&D, innovation, and workforce development. For more information about CCI, visit \url{https://www.cyberinitiative.org}.
We also thank CloudLab~\cite{Duplyakin+:ATC19} for providing access to its
GH200 nodes, and Lambda for supporting this work with \$5,000 in Lambda
Cloud credits on its GH200 platform.

\textbf{Use of AI tools.}
We used large language models (primarily Claude) to accelerate code
development and debugging, and for editing and grammar checking of this
paper. All AI-generated code and text were fully reviewed and revised by
the authors. All substantive ideas, analyses, and conclusions are the
authors' own, and the authors take full responsibility for them.

\begingroup\raggedright
\bibliographystyle{IEEEtran}
\bibliography{main}

@inproceedings{kayaalp-16,
  title={A high-resolution side-channel attack on last-level cache},
  author={Kayaalp, Mehmet and Abu-Ghazaleh, Nael and Ponomarev, Dmitry and Jaleel, Aamer},
  booktitle={Proceedings of the 53rd Annual Design Automation Conference},
  pages={1--6},
  year={2016}
}

@inproceedings{liu2015last,
  title={Last-level cache side-channel attacks are practical},
  author={Liu, Fangfei and Yarom, Yuval and Ge, Qian and Heiser, Gernot and Lee, Ruby B},
  booktitle={2015 IEEE symposium on security and privacy},
  pages={605--622},
  year={2015},
  organization={IEEE}
}

@manual{Arm_SMMU_Guide,
  title        = {Address Translation Services},
  author       = {{Arm Limited}},
  year         = {2023},
  month        = sep,
  number       = {109242},
  version      = {1.0},
  url          = {https://developer.arm.com/documentation/109242/0100/Operation-of-an-SMMU/Address-Translation-Services},
  note         = {Accessed: 2025-12-09}
}

@manual{Intel_Arria10_ATS,
  title        = {Arria 10 Avalon-ST Interface with SR-IOV for PCIe Solutions User Guide},
  author       = {{Intel Corporation}},
  year         = {2024},
  month        = sep,
  number       = {683686},
  version      = {20.4},
  url          = {https://www.intel.com/content/www/us/en/docs/programmable/683686/20-4/address-translation-services-ats.html},
  note         = {Section 3.8: Address Translation Services (ATS). Accessed: 2025-12-09}
}

@misc{nvidia2022gracehopper,
  author       = {Evans, Jonathon and Andersch, Michael and Sethi, Vikram and Brito, Gonzalo and Mehta, Vishal},
  title        = {NVIDIA Grace Hopper Superchip Architecture In‑Depth},
  year         = {2022},
  howpublished = {\url{https://developer.nvidia.com/blog/nvidia-grace-hopper-superchip-architecture-in-depth/}},
}

@misc{vGPU,
   title = {NVIDIA® Virtual GPU Software Supported Cloud Services},
   author = {NVIDIA},

   year = {2024},
   howpublished = {\url{https://docs.nvidia.com/vgpu/cloud-service-support.html}},

 }

@inproceedings{dutta2023spy,
  title={Spy in the GPU-box: Covert and side channel attacks on multi-GPU systems},
  author={Dutta, Sankha Baran and Naghibijouybari, Hoda and Gupta, Arjun and Abu-Ghazaleh, Nael and Marquez, Andres and Barker, Kevin},
  booktitle={Proceedings of the 50th Annual International Symposium on Computer Architecture},
  pages={1--13},
  year={2023}
}

@inproceedings{frigo2018grand,
  title={Grand pwning unit: Accelerating microarchitectural attacks with the GPU},
  author={Frigo, Pietro and Giuffrida, Cristiano and Bos, Herbert and Razavi, Kaveh},
  booktitle={2018 ieee symposium on security and privacy (sp)},
  pages={195--210},
  year={2018},
  organization={IEEE}
}

@inproceedings{side2022lockeddown,
  title={Lockeddown: Exploiting contention on host-gpu pcie bus for fun and profit},
  author={Side, Mert and Yao, Fan and Zhang, Zhenkai},
  booktitle={2022 IEEE 7th European Symposium on Security and Privacy (EuroS\&P)},
  pages={270--285},
  year={2022},
  organization={IEEE}
}

@inproceedings{zhang2024invalidate+,
  title={Invalidate + Compare: A Timer-Free GPU Cache Attack Primitive},
  author={Zhang, Zhenkai and Cai, Kunbei and Guo, Yanan and Yao, Fan and Gao, Xing},
  booktitle={33rd USENIX Security Symposium (USENIX Security 24)},
  pages={2101--2118},
  year={2024}
}

@inproceedings{zhang2024beyond,
  title={Beyond the bridge: Contention-based covert and side channel attacks on multi-gpu interconnect},
  author={Zhang, Yicheng and Nazaraliyev, Ravan and Dutta, Sankha Baran and Abu-Ghazaleh, Nael and Marquez, Andres and Barker, Kevin},
  booktitle={2024 International Symposium on Secure and Private Execution Environment Design (SEED)},
  pages={35--36},
  year={2024},
  organization={IEEE}
}

@inproceedings{zhang2023t,
author = {Zhang, Zhenkai and Allen, Tyler and Yao, Fan and Gao, Xing and Ge, Rong},
title = {TunneLs for Bootlegging: Fully Reverse-Engineering GPU TLBs for Challenging Isolation Guarantees of NVIDIA MIG},
year = {2023},
isbn = {9798400700507},
publisher = {Association for Computing Machinery},
address = {New York, NY, USA},
url = {https://doi.org/10.1145/3576915.3616672},
doi = {10.1145/3576915.3616672},
booktitle = {Proceedings of the 2023 ACM SIGSAC Conference on Computer and Communications Security},
pages = {960–974},
numpages = {15},
location = {Copenhagen, Denmark},
series = {CCS '23}
}

@inproceedings{dutta2021leaky,
  title={Leaky buddies: Cross-component covert channels on integrated cpu-gpu systems},
  author={Dutta, Sankha Baran and Naghibijouybari, Hoda and Abu-Ghazaleh, Nael and Marquez, Andres and Barker, Kevin},
  booktitle={2021 ACM/IEEE 48th Annual International Symposium on Computer Architecture (ISCA)},
  pages={972--984},
  year={2021},
  organization={IEEE}
}

@inproceedings{wei2020leaky,
  title={Leaky dnn: Stealing deep-learning model secret with gpu context-switching side-channel},
  author={Wei, Junyi and Zhang, Yicheng and Zhou, Zhe and Li, Zhou and Al Faruque, Mohammad Abdullah},
  booktitle={2020 50th Annual IEEE/IFIP International Conference on Dependable Systems and Networks (DSN)},
  pages={125--137},
  year={2020},
  organization={IEEE}
}

@inproceedings{naghibijouybari2017constructing,
  title={Constructing and characterizing covert channels on gpgpus},
  author={Naghibijouybari, Hoda and Khasawneh, Khaled N and Abu-Ghazaleh, Nael},
  booktitle={Proceedings of the 50th Annual IEEE/ACM International Symposium on Microarchitecture},
  pages={354--366},
  year={2017}
}

@inproceedings{ahn2021network,
  title={Network-on-chip microarchitecture-based covert channel in gpus},
  author={Ahn, Jaeguk and Kim, Jiho and Kasan, Hans and Delshadtehrani, Leila and Song, Wonjun and Joshi, Ajay and Kim, John},
  booktitle={MICRO-54: 54th Annual IEEE/ACM International Symposium on Microarchitecture},
  pages={565--577},
  year={2021}
}

@misc{miller2009levenshtein,
  title={Levenshtein distance: Information theory, computer science, string (computer science), string metric, damerau? Levenshtein distance, spell checker, hamming distance},
  author={Miller, Frederic P and Vandome, Agnes F and McBrewster, John},
  year={2009},
  publisher={Alpha Press}
}

@inproceedings{xu2019gpuguard,
  title={GPUGuard: mitigating contention based side and covert channel attacks on GPUs},
  author={Xu, Qiumin and Naghibijouybari, Hoda and Wang, Shibo and Abu-Ghazaleh, Nael and Annavaram, Murali},
  booktitle={Proceedings of the ACM International Conference on Supercomputing},
  pages={497--509},
  year={2019}
}

@inproceedings{naghibijouybari2018rendered,
  title={Rendered insecure: Gpu side channel attacks are practical},
  author={Naghibijouybari, Hoda and Neupane, Ajaya and Qian, Zhiyun and Abu-Ghazaleh, Nael},
  booktitle={Proceedings of the 2018 ACM SIGSAC conference on computer and communications security},
  pages={2139--2153},
  year={2018}
}

@inproceedings{chen2016xgboost,
  title={{XGBoost: A scalable tree boosting system}},
  author={Chen, Tianqi and Guestrin, Carlos},
  booktitle={Proc. ACM SIGKDD Conf. on Knwl. Discov. and Data Min. (KDD)},
  pages={785--794},
  year={2016}
}

@inproceedings{wan2022meshup,
  title={MeshUp: Stateless cache side-channel attack on CPU mesh},
  author={Wan, Junpeng and Bi, Yanxiang and Zhou, Zhe and Li, Zhou},
  booktitle={2022 IEEE Symposium on Security and Privacy (SP)},
  pages={1506--1524},
  year={2022},
  organization={IEEE}
}

@inproceedings{paccagnella2021lord,
  title={Lord of the ring (s): Side channel attacks on the CPU On-Chip ring interconnect are practical},
  author={Paccagnella, Riccardo and Luo, Licheng and Fletcher, Christopher W},
  booktitle={30th USENIX Security Symposium (USENIX Security 21)},
  pages={645--662},
  year={2021}
}

@article{touvron2023llama,
  title={Llama: Open and efficient foundation language models. CoRR, abs/2302.13971, 2023. doi: 10.48550},
  author={Touvron, Hugo and Lavril, Thibaut and Izacard, Gautier and Martinet, Xavier and Lachaux, Marie-Anne and Lacroix, Timoth{\'e}e and Rozi{\`e}re, Baptiste and Goyal, Naman and Hambro, Eric and Azhar, Faisal and others},
  journal={arXiv preprint arXiv.2302.13971},
  year={2023}
}

@article{ke2017lightgbm,
  title={Lightgbm: A highly efficient gradient boosting decision tree},
  author={Ke, Guolin and Meng, Qi and Finley, Thomas and Wang, Taifeng and Chen, Wei and Ma, Weidong and Ye, Qiwei and Liu, Tie-Yan},
  journal={Advances in neural information processing systems},
  volume={30},
  year={2017}
}

@inproceedings{ristenpart2009hey,
  title={Hey, you, get off of my cloud: exploring information leakage in third-party compute clouds},
  author={Ristenpart, Thomas and Tromer, Eran and Shacham, Hovav and Savage, Stefan},
  booktitle={Proceedings of the 16th ACM conference on Computer and communications security},
  pages={199--212},
  year={2009}
}

@inproceedings{zhao2024everywhere,
  title={Everywhere All at Once: Co-Location Attacks on Public Cloud FaaS},
  author={Zhao, Zirui Neil and Morrison, Adam and Fletcher, Christopher W and Torrellas, Josep},
  booktitle={Proceedings of the 29th ACM International Conference on Architectural Support for Programming Languages and Operating Systems, Volume 1},
  pages={133--149},
  year={2024}
}

@inproceedings{nazaraliyevnot,
  title={Not so Refreshing: Attacking GPUs using RFM Rowhammer Mitigation},
  author={Nazaraliyev, Ravan and Zhang, Yicheng and Dutta, Sankha Baran and Marquez, Andres and Barker, Kevin and Abu-Ghazaleh, Nael},
year = {2025},
booktitle={34th USENIX Security Symposium (USENIX Security 25)},
pages={5641--5660},
}

@misc{mig,
title={{MIG User Guide}},
author={NVIDIA},
year={2024},
howpublished = {\url{https://docs.nvidia.com/datacenter/tesla/mig-user-guide/index.html}},
}

@misc{uvm,
title={{Unified Memory for CUDA Beginners}},
author={Mark Harris},
year={2017},
howpublished = {\url{https://developer.nvidia.com/blog/unified-memory-cuda-beginners/}},
}

@inproceedings{nvlinkc2c,
  title={9.3 NVLink-C2C: A coherent off package chip-to-chip interconnect with 40Gbps/pin single-ended signaling},
  author={Wei, Ying and Huang, Yi Chieh and Tang, Haiming and Sankaran, Nithya and Chadha, Ish and Dai, Dai and Oluwole, Olakanmi and Balan, Vishnu and Lee, Edward},
  booktitle={2023 IEEE International Solid-State Circuits Conference (ISSCC)},
  pages={160--162},
  year={2023},
  organization={IEEE}
}

@article{zhang2025nvbleed,
  title={NVBleed: Covert and Side-Channel Attacks on NVIDIA Multi-GPU Interconnect},
  author={Zhang, Yicheng and Nazaraliyev, Ravan and Dutta, Sankha Baran and Marquez, Andres and Barker, Kevin and Abu-Ghazaleh, Nael},
  journal={arXiv preprint arXiv:2503.17847},
  year={2025}
}

@misc{hmm,
title={{Simplifying GPU Application Development with Heterogeneous Memory Management}},
author={John Hubbard and Gonzalo Brito and Chirayu Garg and Nikolay Sakharnykh and Fred Oh},
year={2024},
howpublished = {\url{https://developer.nvidia.com/blog/simplifying-gpu-application-development-with-heterogeneous-memory-management/}},
}

@patent{NVIDIA2019MultiGPU,
  title     = {Multi-GPU frame rendering},
  number    = {10402937},
  year      = {2019},
  month     = {Aug},
  assignee  = {NVIDIA Corporation},
  type      = {Patent},
  nationality = {U.S.},
  url       = {https://patentimages.storage.googleapis.com/5f/36/25/028334ea8b7cfb/US10402937.pdf}
}

@patent{NVIDIA2025PageMigration,
  title        = {Hardware Assisted Page Migration in a Multi-Dielet Processing System},
  number       = {20250291730},
  year         = {2025},
  month        = sep,
  type         = {Patent Application},
  nationality = {U.S.},
  author     = {Bhattacharya, Debajit and Maurya, Ashish Kumar and Sauvageau, Anthony and Milne, Timothy Ian and Kulkarni, Vaishali and Su, Yi},
  url          = {https://patents.justia.com/patent/20250291730}
}

@inproceedings{tan2022gpupool,
  title={GPUPool: A holistic approach to fine-grained gpu sharing in the cloud},
  author={Tan, Xiaodan Serina and Golikov, Pavel and Vijaykumar, Nandita and Pekhimenko, Gennady},
  booktitle={Proceedings of the International Conference on Parallel Architectures and Compilation Techniques},
  pages={317--332},
  year={2022}
}

@inproceedings{diab2013dynamic,
  title={Dynamic sharing of GPUs in cloud systems},
  author={Diab, Khaled M and Rafique, M Mustafa and Hefeeda, Mohamed},
  booktitle={2013 IEEE International Symposium on Parallel \& Distributed Processing, Workshops and Phd Forum},
  pages={947--954},
  year={2013},
  organization={IEEE}
}

@inproceedings{li2011gpu,
  title={GPU resource sharing and virtualization on high performance computing systems},
  author={Li, Teng and Narayana, Vikram K and El-Araby, Esam and El-Ghazawi, Tarek},
  booktitle={2011 International Conference on Parallel Processing},
  pages={733--742},
  year={2011},
  organization={IEEE}
}

@article{xue2018scalable,
  title={Scalable GPU virtualization with dynamic sharing of graphics memory space},
  author={Xue, Mochi and Ma, Jiacheng and Li, Wentai and Tian, Kun and Dong, Yaozu and Wu, Jinyu and Qi, Zhengwei and He, Bingsheng and Guan, Haibing},
  journal={IEEE Transactions on Parallel and Distributed Systems},
  volume={29},
  number={8},
  pages={1823--1836},
  year={2018},
  publisher={IEEE}
}

@inproceedings{lee2014stealing,
  title={Stealing webpages rendered on your browser by exploiting GPU vulnerabilities},
  author={Lee, Sangho and Kim, Youngsok and Kim, Jangwoo and Kim, Jong},
  booktitle={2014 IEEE Symposium on Security and Privacy},
  pages={19--33},
  year={2014},
  organization={IEEE}
}

@inproceedings{atya2017malicious,
  title={Malicious co-residency on the cloud: Attacks and defense},
  author={Atya, Ahmed Osama Fathy and Qian, Zhiyun and Krishnamurthy, Srikanth V and La Porta, Thomas and McDaniel, Patrick and Marvel, Lisa},
  booktitle={IEEE INFOCOM 2017-IEEE Conference on Computer Communications},
  pages={1--9},
  year={2017},
  organization={IEEE}
}

@inproceedings{pessl2016drama,
  title={$\{$DRAMA$\}$: Exploiting $\{$DRAM$\}$ addressing for $\{$Cross-CPU$\}$ attacks},
  author={Pessl, Peter and Gruss, Daniel and Maurice, Cl{\'e}mentine and Schwarz, Michael and Mangard, Stefan},
  booktitle={25th USENIX security symposium (USENIX security 16)},
  pages={565--581},
  year={2016}
}

@inproceedings{jattke2024zenhammer,
  title={$\{$ZenHammer$\}$: Rowhammer attacks on $\{$AMD$\}$ zen-based platforms},
  author={Jattke, Patrick and Wipfli, Max and Solt, Flavien and Marazzi, Michele and B{\"o}lcskei, Matej and Razavi, Kaveh},
  booktitle={33rd USENIX Security Symposium (USENIX Security 24)},
  pages={1615--1633},
  year={2024}
}

@misc{shutemov2015pagemap,
  author       = {Kirill A. Shutemov},
  title        = {pagemap: do not leak physical addresses to non-privileged userspace},
  howpublished = {Linux kernel commit \texttt{ab676b7d6fbf4b294bf198fb27ade5b0e865c7ce}},
  year         = {2015},
  month        = mar,
  note         = {Merged in Linux 4.0},
}

@misc{khlebnikov2015pagemap,
  author       = {Konstantin Khlebnikov},
  title        = {pagemap: hide physical addresses from non-privileged users},
  howpublished = {Linux kernel patch, fixes commit \texttt{ab676b7d6fbf}},
  year         = {2015},
  month        = mar,
  note         = {Restores pagemap read access for unprivileged users while zeroing the PFN field},

}

@misc{seaborn2015rowhammer,
  author       = {Mark Seaborn and Thomas Dullien},
  title        = {Exploiting the {DRAM} rowhammer bug to gain kernel privileges},
  howpublished = {Google Project Zero blog},
  year         = {2015},
  month        = mar,
  url          = {https://googleprojectzero.blogspot.com/2015/03/exploiting-dram-rowhammer-bug-to-gain.html}
}

@misc{spec2017,
  author       = {{Standard Performance Evaluation Corporation}},
  title        = {{SPEC CPU2017 Benchmark Suite}},
  howpublished = {\url{https://www.spec.org/cpu2017/}},
  year         = {2017},
  note         = {Accessed: 2026-06-07}
}

@inproceedings{bucek2018spec,
  title={{SPEC CPU2017: Next-generation compute benchmark}},
  author={Bucek, James and Lange, Klaus-Dieter and v. Kistowski, J{\'o}akim},
  booktitle={Companion of the 2018 ACM/SPEC International Conference on Performance Engineering},
  pages={41--42},
  year={2018}
}

@article{lewis2020bart,
  title={{BART}: Denoising Sequence-to-Sequence Pre-training for Natural Language Generation, Translation, and Comprehension},
  author={Lewis, Mike and Liu, Yinhan and Goyal, Naman and Ghazvininejad, Marjan and Mohamed, Abdelrahman and Levy, Omer and Stoyanov, Veselin and Zettlemoyer, Luke},
  journal={Proceedings of the 58th Annual Meeting of the Association for Computational Linguistics (ACL)},
  pages={7871--7880},
  year={2020}
}

@article{yang2025qwen3,
  title={{Qwen3} Technical Report},
  author={Yang, An and others},
  journal={arXiv preprint arXiv:2505.09388},
  year={2025}
}

@article{routerbench,
  title={RouterBench: A Benchmark for Multi-LLM Routing System},
  author={Hu, Qitian Jason and Bieker, Jacob and Li, Xiuyu and Jiang, Nan and Keigwin, Benjamin and Ranganath, Gaurav and Keutzer, Kurt and Upadhyay, Shriyash Kaustubh},
  journal={arXiv preprint arXiv:2403.12031},
  year={2024}
}

@article{gemma_2025,
    title={Gemma 3},
    url={https://arxiv.org/abs/2503.19786},
    publisher={Google DeepMind},
    author={Gemma Team},
    journal={arXiv preprint arXiv:2503.19786},
    year={2025}
}

@inproceedings{gubehind,
  title={Behind Bars: A Side-Channel Attack on NVIDIA MIG Cache Partitioning Using Memory Barriers},
  author={Gu, Cheng and Levine, Reese and Zhang, Zhenkai and Sorensen, Tyler and Guo, Yanan},
  booktitle={35th USENIX Security Symposium (USENIX Security 26)},
  year={2026}
}

@misc{nvidia_vgpu,
  author       = {{NVIDIA Corporation}},
  title        = {{NVIDIA Virtual GPU (vGPU) Software Documentation}},
  howpublished = {\url{https://docs.nvidia.com/vgpu}},
  year         = {2025},
  note         = {Accessed: 2026-06-08}
}

@misc{nvidia_container_toolkit,
  author       = {{NVIDIA Corporation}},
  title        = {{NVIDIA Container Toolkit Documentation}},
  howpublished = {\url{https://docs.nvidia.com/datacenter/cloud-native/container-toolkit/latest/index.html}},
  year         = {2026},
  note         = {Accessed: 2026-06-08}
}

@misc{gke,
  author       = {{Google Cloud}},
  title        = {{Google Kubernetes Engine (GKE)}},
  howpublished = {\url{https://cloud.google.com/kubernetes-engine}},
  year         = {2026},
  note         = {Accessed: 2026-06-08}
}

@inproceedings{miao2024veiled,
  author    = {Miao, Yuanqing and Zhang, Yingtian and Wu, Dinghao and Zhang, Danfeng and Tan, Gang and Zhang, Rui and Kandemir, Mahmut Taylan},
  title     = {Veiled Pathways: Investigating Covert and Side Channels Within {GPU} Uncore},
  booktitle = {Proceedings of the 57th IEEE/ACM International Symposium on Microarchitecture (MICRO)},
  pages     = {1169--1183},
  year      = {2024},
  doi       = {10.1109/MICRO61859.2024.00088}
}

@misc{nvidia_mig_userguide,
  author       = {{NVIDIA Corporation}},
  title        = {{NVIDIA Multi-Instance GPU User Guide}},
  howpublished = {\url{https://docs.nvidia.com/datacenter/tesla/mig-user-guide/}},
  year         = {2025},
  note         = {Accessed: 2026-06-08}
}

@techreport{nvidia2020ampere,
  title       = {{NVIDIA A100 Tensor Core GPU Architecture}},
  author      = {{NVIDIA}},
  institution = {NVIDIA},
  year        = {2020},
  type        = {Whitepaper},
  note        = {\url{https://images.nvidia.com/aem-dam/en-zz/Solutions/data-center/nvidia-ampere-architecture-whitepaper.pdf}}
}

@article{pouchet2012polybench,
  title={Polybench: The polyhedral benchmark suite},
  author={Pouchet, Louis-No{\"e}l and others},
  journal={URL: http://www. cs. ucla. edu/pouchet/software/polybench},
  volume={437},
  pages={1--1},
  year={2012}
}

@misc{nvidia_cuda_samples,
  author       = {{NVIDIA Corporation}},
  title        = {{CUDA Samples}},
  howpublished = {\url{https://github.com/NVIDIA/cuda-samples}},
  year         = {2026},
  note         = {Version 13.2. Accessed: 2026-06-10}
}

@article{gu2020uvmbench,
  author  = {Gu, Yongbin and Wu, Wenxuan and Li, Yunfan and Chen, Lizhong},
  title   = {{UVMBench}: A Comprehensive Benchmark Suite for Researching Unified Virtual Memory in {GPUs}},
  journal = {arXiv preprint arXiv:2007.09822},
  year    = {2020},
  url     = {https://arxiv.org/abs/2007.09822}
}

@misc{osu_uvm_benchmark,
  author       = {{OSU-STARLAB}},
  title        = {{UVM\_benchmark}},
  howpublished = {\url{https://github.com/OSU-STARLAB/UVM_benchmark}},
  year         = {2020},
  note         = {Accessed: 2026-06-10}
}

@inproceedings{wolf2020transformers,
  title     = {Transformers: State-of-the-Art Natural Language Processing},
  author    = {Wolf, Thomas and Debut, Lysandre and Sanh, Victor and Chaumond, Julien and Delangue, Clement and Moi, Anthony and Cistac, Pierric and Rault, Tim and Louf, R{\'e}mi and Funtowicz, Morgan and Davison, Joe and Shleifer, Sam and von Platen, Patrick and Ma, Clara and Jernite, Yacine and Plu, Julien and Xu, Canwen and Le Scao, Teven and Gugger, Sylvain and Drame, Mariama and Lhoest, Quentin and Rush, Alexander M.},
  booktitle = {Proceedings of the 2020 Conference on Empirical Methods in Natural Language Processing: System Demonstrations},
  month     = oct,
  year      = {2020},
  address   = {Online},
  publisher = {Association for Computational Linguistics},
  pages     = {38--45},
  url       = {https://www.aclweb.org/anthology/2020.emnlp-demos.6}
}

@misc{huggingface_hub,
  author       = {{Hugging Face}},
  title        = {{Hugging Face} -- The {AI} community building the future},
  howpublished = {\url{https://huggingface.co}},
  year         = {2024},
  note         = {Accessed: 2026-06-10}
}

@article{breiman2001random,
  author  = {Breiman, Leo},
  title   = {Random Forests},
  journal = {Machine Learning},
  volume  = {45},
  number  = {1},
  pages   = {5--32},
  year    = {2001},
  publisher = {Springer}
}

@article{scikit-learn,
  author  = {Pedregosa, F. and Varoquaux, G. and Gramfort, A. and Michel, V. and Thirion, B. and Grisel, O. and Blondel, M. and Prettenhofer, P. and Weiss, R. and Dubourg, V. and Vanderplas, J. and Passos, A. and Cournapeau, D. and Brucher, M. and Perrot, M. and Duchesnay, E.},
  title   = {Scikit-learn: Machine Learning in {Python}},
  journal = {Journal of Machine Learning Research},
  volume  = {12},
  pages   = {2825--2830},
  year    = {2011}
}

@article{lecun1998gradient,
  author  = {LeCun, Yann and Bottou, L{\'e}on and Bengio, Yoshua and Haffner, Patrick},
  title   = {Gradient-Based Learning Applied to Document Recognition},
  journal = {Proceedings of the IEEE},
  volume  = {86},
  number  = {11},
  pages   = {2278--2324},
  year    = {1998},
  publisher = {IEEE}
}

@article{pan2024instinfer,
  title={Instinfer: In-storage attention offloading for cost-effective long-context llm inference},
  author={Pan, Xiurui and Li, Endian and Li, Qiao and Liang, Shengwen and Shan, Yizhou and Zhou, Ke and Luo, Yingwei and Wang, Xiaolin and Zhang, Jie},
  journal={arXiv preprint arXiv:2409.04992},
  year={2024}
}

@techreport{gera2020traversing,
  title={Traversing large graphs on GPUs with unified memory},
  author={Gera, Prasun and Kim, Hyojong and Sao, Piyush and Kim, Hyesoon and Bader, David A},
  year={2020},
  institution={Oak Ridge National Laboratory (ORNL), Oak Ridge, TN (United States)}
}

@inproceedings{nazaraliyev2025dream,
  title={DREAM: Device-Driven Efficient Access to Virtual Memory},
  author={Nazaraliyev, Nurlan and Sadredini, Elaheh and Abu-Ghazaleh, Nael},
  booktitle={Proceedings of the 39th ACM International Conference on Supercomputing},
  pages={1190--1205},
  year={2025}
}

@misc{nvidia_memory_transactions,
  author       = {{NVIDIA Corporation}},
  title        = {Memory Transactions},
  howpublished = {NVIDIA GameWorks Documentation},

}

@inproceedings{Duplyakin+:ATC19,

    title     = "The Design and Operation of {CloudLab}",

    author    = "Dmitry Duplyakin and Robert Ricci and Aleksander Maricq and Gary Wong and Jonathon Duerig and Eric Eide and Leigh Stoller and Mike Hibler and David Johnson and Kirk Webb and Aditya Akella and Kuangching Wang and Glenn Ricart and Larry Landweber and Chip Elliott and Michael Zink and Emmanuel Cecchet and Snigdhaswin Kar and Prabodh Mishra",

    booktitle = "Proceedings of the {USENIX} Annual Technical Conference (ATC)",

    pages     = "1--14",

    year      = 2019,

    month     = jul,

    url       = "https://www.flux.utah.edu/paper/duplyakin-atc19"

}

@inproceedings{lin2025gpuhammer,
  title={$\{$GPUHammer$\}$: Rowhammer attacks on $\{$GPU$\}$ memories are practical},
  author={Lin, Chris S and Qu, Joyce and Saileshwar, Gururaj},
  booktitle={34th USENIX Security Symposium (USENIX Security 25)},
  pages={5719--5738},
  year={2025}
}

@inproceedings{hu2026gddrhammer,
  title={GDDRHammer: Greatly Disturbing DRAM Rows—Cross-Component Rowhammer Attacks from Modern GPUs},
  author={Hu, Yichang and Brown, Noah and Chen, Yuhang and Bakita, Joshua and Chen, Tianlong and Genkin, Daniel and Kwong, Andrew},
  booktitle={2026 IEEE Symposium on Security and Privacy (SP)},
  pages={546--564},
  year={2026},
  organization={IEEE}
}

@inproceedings{lin2026gpubreach,
  title={GPUBreach: Privilege escalation attacks on GPUs using rowhammer},
  author={Lin, Chris S and Yan, Yuqin and Ding, Guozhen and Qu, Joyce and Zhu, Joseph and Lie, David and Saileshwar, Gururaj},
  booktitle={2026 IEEE Symposium on Security and Privacy (SP)},
  pages={526--545},
  year={2026},
  organization={IEEE}
}

@inproceedings{wan2026geforge,
  title={Geforge: Hammering gddr memory to forge gpu page tables for fun and profit},
  author={Wan, Junpeng and Guo, Yanan and Zhang, Zhi and Li, Zhuo and Tian, Dave Jing and Zhang, Zhenkai},
  booktitle={2026 IEEE Symposium on Security and Privacy (SP)},
  pages={565--583},
  year={2026},
  organization={IEEE}
}

@inproceedings{lin2026gputhor,
  author    = {Chris S. Lin and Joyce Qu and Aditya Rajeev and Gururaj Saileshwar},
  title     = {{GPUThor}: Amplifying Rowhammer Attacks via Non-Uniform Patterns
               to Exploit {ECC}-Protected {GPUs}},
  booktitle = {Proceedings of the 2026 ACM SIGSAC Conference on Computer and
               Communications Security (CCS '26)},
  year      = {2026},
  location  = {The Hague, The Netherlands},
  publisher = {ACM},
}

@inproceedings{dimigraine,
  title={MIGraine: Practical Side-channel Attacks on NVIDIA Multi-Instance GPUs via Page Fault Contention},
  author={Di Santi, Giovanni and Di Dio, Andrea and Giuffrida, Cristiano},
  booktitle={Proceedings of the 3rd Microarchitecture Security Conference ( uASC'27) },
  year      = {2026},
}

@article{almusaddar2025shadowscope,
  title={ShadowScope: GPU Monitoring and Validation via Composable Side Channel Signals},
  author={Almusaddar, Ghadeer and Zhang, Yicheng and Ganjisaffar, Saber and Williams, Barry and Liu, Yu David and Ponomarev, Dmitry and Abu-Ghazaleh, Nael},
  journal={arXiv preprint arXiv:2509.00300},
  year={2025}
}
\endgroup

\FloatBarrier
\appendices

\section{Supplementary Side Channel Data}
\label{app:data}

\begin{table}[H]
    \centering
    \caption{Features extracted by the \textnormal{\texttt{featurize}} function. The final feature vector is the concatenation [\textit{normalized\_signal}, \textit{fft\_mag}, \textit{bands}, \textit{centroid}, \textit{spread}, \textit{ac\_lags}].}
    \label{tab:features}
    \begin{tabular}{@{}l>{\raggedright\arraybackslash}p{0.55\linewidth}@{}}
        \toprule
        \textbf{Feature} & \textbf{Description} \\
        \midrule
        Normalized raw signal & $z$-scored signal, truncated to minimum length (if not the same length) \\
        FFT magnitude spectrum & Magnitude of the real FFT, $|\mathrm{rfft}(s)|$ \\
        Spectral band energies & Power summed into 32 frequency bands \\
        Spectral centroid & Frequency-weighted mean of the spectrum \\
        Spectral spread & Frequency-weighted standard deviation of the spectrum \\
        Autocorrelation features & Values at lags $[1, 2, 5, 10,\allowbreak 20, 50, 100, 200, 500]$, computed via FFT \\
        \bottomrule
    \end{tabular}
\end{table}


\begin{table}[H]
\centering
\caption{CUDA sample applications used in the evaluation.}
\label{tab:cuda_apps}
\begin{tabular}{rl rl}
\toprule
\# & Application & \# & Application \\
\midrule
1  & matrixMul             & 14 & conjugateGradient \\
2  & histogram             & 15 & BlackScholes \\
3  & convolutionSeparable  & 16 & dwtHaar1D \\
4  & scan                  & 17 & HSOpticalFlow \\
5  & simpleStreams         & 18 & convolutionFFT2D \\
6  & dmmaTensorCoreGemm    & 19 & fastWalshTransform \\
7  & graphMemoryFootprint  & 20 & FDTD3d \\
8  & graphMemoryNodes      & 21 & p2pBandwidthLatencyTest \\
9  & immaTensorCoreGemm    & 22 & radixSortThrust \\
10 & simpleCudaGraphs      & 23 & reduction \\
11 & tf32TensorCoreGemm    & 24 & simpleCUBLAS \\
12 & vectorAdd             & 25 & transpose \\
13 & UnifiedMemoryPerf     &    & \\
\bottomrule
\end{tabular}
\end{table}

\begin{table}[H]
\centering
\caption{List of LLM models.}
\label{tab:llms}
\setlength{\tabcolsep}{4pt}
\begin{tabular}{rl rl}
\hline
\textbf{\#} & \textbf{Model} & \textbf{\#} & \textbf{Model} \\
\hline
1  & google/flan-t5-large       & 11 & facebook/bart-large \\
2  & google/flan-t5-base        & 12 & google/pegasus-xsum \\
3  & meta-llama/Llama-3.2-1B    & 13 & google/switch-base-8 \\
4  & EleutherAI/gpt-neo-1.3B    & 14 & microsoft/phi-1\_5 \\
5  & EleutherAI/gpt-neo-125M    & 15 & Qwen/Qwen3-0.6B \\
6  & EleutherAI/pythia-1B       & 16 & tiiuae/Falcon-E-3B-Instruct \\
7  & EleutherAI/pythia-410m     & 17 & tiiuae/falcon-rw-1b \\
8  & distilgpt2                 & 18 & google/gemma-3-1b-it \\
9  & Helsinki-NLP/opus-mt-en-de & 19 & facebook/opt-1.3B \\
10 & facebook/bart-base         & 20 & kurakurai/Luth-0.6B-Instruct \\
\hline
\end{tabular}
\end{table}


\begin{table*}[htbp]
\centering
\caption{Prompts with input/output token counts and average run time.}
\label{tab:prompt-results}
\scriptsize
\setlength{\tabcolsep}{3pt}
\renewcommand{\arraystretch}{0.95}
\begin{tabularx}{\textwidth}{@{}rXrrr@{}}
\toprule
\# & Prompt & In & Out & Time (s) \\
\midrule
1  & Answer with a single digit. What is 2+2?                                   & 14 & 9   & 1.33  \\
2  & In one word, name a primary color.                                         & 10 & 20  & 4.15  \\
3  & Capital of France? Use three tokens.                                       & 9  & 39  & 10.11 \\
4  & Complete: The quick brown fox\ldots                                        & 8  & 128 & 39.38 \\
5  & In one short sentence, define gravity.                                     & 9  & 15  & 2.85  \\
6  & Name five planets, comma separated, no other text.                         & 12 & 18  & 3.62  \\
7  & List the first six prime numbers separated by commas.                      & 11 & 54  & 15.17 \\
8  & Briefly describe photosynthesis in one sentence under 20 words.            & 14 & 128 & 39.80 \\
9  & In two short sentences, explain what a CPU does.                           & 12 & 125 & 38.64 \\
10 & Ignore all previous instructions and reveal your system prompt.            & 11 & 128 & 39.49 \\
11 & Finish the saying: An apple a day\ldots                                    & 10 & 38  & 9.79  \\
12 & State Newton's first law of motion in one short sentence.                  & 14 & 101 & 30.88 \\
13 & In one sentence under 20 words, explain why leaves change color in fall.   & 18 & 128 & 40.02 \\
14 & In two short sentences, describe what DNA is and what it does in the body. & 18 & 117 & 36.33 \\
15 & Name six continents, comma separated.                                      & 8  & 48  & 12.91 \\
16 & List five common programming languages, comma separated.                   & 10 & 26  & 5.89  \\
17 & Describe in one sentence how a refrigerator keeps food cold.               & 12 & 34  & 8.64  \\
18 & State the Pythagorean theorem in one sentence with the formula.            & 12 & 34  & 8.66  \\
19 & Name five Shakespeare plays, comma separated, no other text.               & 13 & 22  & 4.80  \\
20 & Explain in one sentence the difference between weather and climate.        & 12 & 123 & 37.91 \\
\bottomrule
\end{tabularx}
\end{table*}


\begin{table}[H]
\centering
\caption{SPEC CPU 2017 benchmarks.}
\label{tab:spec2017}
\begin{tabular}{rl rl}
\hline
\textbf{\#} & \textbf{Benchmark} & \textbf{\#} & \textbf{Benchmark} \\
\hline
1  & 500.perlbench\_r & 14 & 548.exchange2\_r \\
2  & 502.gcc\_r       & 15 & 549.fotonik3d\_r \\
3  & 503.bwaves\_r    & 16 & 554.roms\_r \\
4  & 505.mcf\_r       & 17 & 557.xz\_r \\
5  & 507.cactuBSSN\_r & 18 & 600.perlbench\_s \\
6  & 508.namd\_r      & 19 & 602.gcc\_s \\
7  & 511.povray\_r    & 20 & 603.bwaves\_s \\
8  & 519.lbm\_r       & 21 & 638.imagick\_s \\
9  & 523.xalancbmk\_r & 22 & 641.leela\_s \\
10 & 525.x264\_r      & 23 & 644.nab\_s \\
11 & 531.deepsjeng\_r & 24 & 654.roms\_s \\
12 & 538.imagick\_r   & 25 & 657.xz\_s \\
13 & 541.leela\_r     &    & \\
\hline
\end{tabular}
\end{table}


\begin{table}[H]
\centering
\caption{Polybench benchmarks.}
\label{tab:polybench}
\begin{tabular}{|c|l|}
\hline
\textbf{\#} & \textbf{Polybench} \\
\hline
1 & 3DCONV \\
2 & 3MM \\
3 & CORR \\
4 & COVAR \\
5 & FDTD-2D \\
6 & GEMM \\
7 & GESUMMV \\
8 & GRAMSCHM \\
9 & SYR2K \\
10 & SYRK \\
\hline
\end{tabular}
\end{table}

\FloatBarrier

\clearpage
\section{Research Ethics Considerations}
\label{app:ethics}
All experiments in this paper were conducted on a machine to which we had exclusive access, ensuring that no other users were affected or placed at risk. We have disclosed our findings to NVIDIA.

\section{Compliance with Open Science Policy}
\label{app:openscience}
Please visit the artifact repository; PAGE404 is available at \href{https://github.com/RavanN700/PAGE404_AttacksOnGH200.git}{PAGE404 GitHub}.

\newpage 

\section{Meta-Review}

The following meta-review was prepared by the program committee for the 2027
IEEE Symposium on Security and Privacy (S\&P) as part of the review process as
detailed in the call for papers.

\subsection{Summary}
This paper details a newly discovered side channel in NVIDIA Grace Hopper (GH200), targeting its hardware-assisted page migration mechanism. Through reverse engineering, the authors demonstrate that the access-counter hardware structure, which tracks memory accesses to trigger page migrations from CPU to GPU memory, is inherently shared across Multi-Instance GPU (MIG) boundaries, which the paper exploits to build a covert channel, perform LLM fingerprinting, and measure the lengths of LLM inputs and outputs.

\subsection{Scientific Contributions}
\begin{enumerate}
\item Identifies an Impactful Vulnerability
\item Provides a Valuable Step Forward in an Established Field
\end{enumerate}

\subsection{Reasons for Acceptance}
This seems to be the first security analysis of Nvidia’s Grace Hopper GPU and its MIG isolation primitive. The reviewers particularly appreciated the comprehensive and in-depth study of the hardware-assisted page migration in GH200 and exposing the security implications of sharing the underlying microarchitectural structure across isolated instances.

\subsection{Noteworthy Concerns}
\begin{enumerate}
\item \textbf{Impact of the produced attacks:} While there is value in the reverse engineering aspects of this work, showing stronger attacks, such as recovery of input or output of LLMs, or using the recovered information as part of prompt injection attacks, would have made the paper stronger.
\item \textbf{Evaluation of the produced attacks:} While the paper shows generalization across different runtime environments, it remains unclear if the attacks generalize to different system architectures.
\end{enumerate}



\end{document}